\documentclass[times,3p,twocolumn]{elsarticle}

\usepackage{amsmath}
\usepackage{amssymb}
\usepackage{hyperref}
\usepackage{multirow}
\usepackage{subfigure}
\usepackage{url}
\usepackage{color}
\usepackage{ulem}

\newcommand{\ex}[1]{\mathrm{e}^{#1}} 
\newcommand{\geant}{\ensuremath{\textsc{Geant4}}} 
\newcommand{\mgt}{\ensuremath{\texttt{mgt}}} 

\journal{Nuclear Instruments and Methods in Physics Research A}

\begin{document}


\begin{frontmatter}
  \title{Discrimination of gamma rays due to inelastic neutron scattering in AGATA}
  \author[au]{A. Ata\c{c}\corref{cor1}}
  \ead{atac@science.ankara.edu.tr}
  \author[au]{A. Ka\c{s}ka\c{s}}
  \author[au]{S. Akkoyun\fnref{fn1}}
  \author[au]{M. \c{S}enyi\u{g}it}
  \author[au]{T. H{\"u}y{\"u}k}
  \author[au]{S. O. Kara}
  \author[uu]{J. Nyberg}
  \ead{johan.nyberg@physics.uu.se} 
  \address[au]{Department of Physics, Faculty of Science, Ankara
    University, 06100 Tando\u{g}an, Ankara, Turkey}
  \address[uu]{Department of Physics and Astronomy, Uppsala
    University, SE-75121 Uppsala, Sweden}
  \cortext[cor1]{Corresponding author}
  \fntext[fn1]{Present address: Cumhuriyet \"Universitesi, 58140,
    Campus Sivas, Turkey}
  \begin{abstract}
    Possibilities of
    discriminating neutrons and $\gamma$ rays in the AGATA
    $\gamma$-ray tracking spectrometer have been investigated with the aim of
    reducing the background due to
    inelastic scattering of neutrons in the high-purity germanium
    crystals. This background may become a 
    serious problem
    especially in experiments with neutron-rich radioactive ion beams.
    Simulations using the {\geant} toolkit and a tracking program based on the forward tracking algorithm 
    were carried out by emitting neutrons and $\gamma$ rays from
    the center of AGATA. Three different methods were developed and tested 
    in order to find
    ``fingerprints'' of the neutron interaction points in the
    detectors. In a simulation with simultaneous emission of six
    neutrons with energies in the range 1-5 MeV and ten $\gamma$ rays
    with energies between 150 and 1450 keV, the peak-to-background
    ratio at a $\gamma$-ray energy of 1.0 MeV was improved
    by a factor
    of 2.4 after neutron rejection with a reduction of the photopeak
    efficiency at 1.0 MeV of only a factor of 1.25.
  \end{abstract}
\end{frontmatter}

\section{Introduction}

One of the large challenges of contemporary experimental nuclear
structure physics is to study exotic nuclides, which lie far from the
line of $\beta$ stability. For production of such proton- and
neutron-rich nuclei the usage of radioactive ion beams is
essential. The intensity of such beams is, however, several orders of
magnitude lower than what is obtainable with stable beams and the
production rates of the exotic nuclides are very low.  In order to
partly compensate for the low production rates highly efficient
detection techniques are being developed. One such technique is
$\gamma$-ray tracking \cite{1999NIMPA.422..195}.  Two $\gamma$-ray
tracking arrays 
made of segmented high-purity germanium (HPGe) detectors
are currently under development, AGATA \cite{AGATA} in
Europe and GRETA \cite{GRETA} in the USA. These future instruments
promise an increase in performance over existing detector arrays by
several orders of magnitude for certain experimental conditions. This
increase in the experimental sensitivity opens up the opportunity to
study very weak reaction channels and interesting nuclear phenomena.

One of the most common types of nuclear reactions used in nuclear
structure physics is the heavy-ion fusion-evaporation (HIFE) reaction
in which two nuclei fuse to form a compound nucleus that decays by
evaporating a number of light particles, mainly neutrons, protons,
and/or $\alpha$ particles. In contrast to the light charged particles,
neutrons can travel long distances and interact within the germanium
detectors together with the $\gamma$ rays emitted by the residual
nuclei.  Neutrons are detected indirectly in AGATA either via elastic
scattering with the Ge nuclei or via nuclear reactions which may
produce neutrons, charged particles and $\gamma$ rays. Elastic
scattering, $^{\textrm{nat}}$Ge(n,n)$^{\textrm{nat}}$Ge, gives the
largest contribution to the total cross-section in the neutron energy
range of interest for HIFE reactions, from about 1 to 10 MeV, while
inelastic scattering,
$^{\textrm{nat}}$Ge(n,n$^{\prime}\gamma$)$^{\textrm{nat}}$Ge, has the
second largest cross section in this energy range
\cite{2005NIMPA.546..553}. The expected probability of neutrons with
energies 1-10 MeV to be detected by AGATA is about 50~\%, if the
low-energy threshold of the Ge detector signals is set to 5 keV
(target value for AGATA) \cite{2005NIMPA.550..379}.

In several recent articles
\cite{2005NIMPA.546..553,2005NIMPA.550..379,2008EURPH...,2009NIM...},
experimental studies of neutron interactions in segmented HPGe
detectors were reported with the aim of reducing the neutron-induced
background in the $\gamma$-ray spectra and to utilize segmented HPGe
detector arrays to measure neutrons. In an earlier Monte Carlo
simulation \cite{2005NIMPA.550..379} of the AGATA $\gamma$-ray
spectrometer, the background in the $\gamma$-ray spectra due to the
inelastic scattering of neutrons was discussed and different
possibilities of background suppression were investigated without any
conclusive results. The neutron background is expected to become a
serious problem in future studies, especially when using neutron-rich
radioactive ion beams of the coming facilities, e.g. NUSTAR/FAIR
\cite{npn_nustar,2007PPNP...59..3}, and SPIRAL2
\cite{2007PPNP...59..22}, and in experiments where a low background in
the $\gamma$-ray spectra is of importance.

In this work, a {\geant} \cite{2003NIMPA.506..250} based Monte Carlo
simulation of the AGATA spectrometer \cite{Farnea2003} has been
performed in order to investigate the possibility of reducing the
neutron induced background by $\gamma$-ray tracking. The tracking
algorithm of the {\mgt} program \cite{2004NP...746..248,Bazzacco-MGT} has
been used to identify the neutron interaction points and methods for
eliminating the $\gamma$ rays originated from inelastic neutron
scattering have been developed with a special emphasis on not reducing
the photopeak (= full-energy peak) efficiency of the $\gamma$ rays of
interest.

An obvious method of distinguishing between neutrons and $\gamma$ rays
is to use the time-of-flight technique. This possibility was
previously investigated in a simulation of the AGATA array
\cite{2005NIMPA.546..553}. It is still an open question if the time
resolution that can be achieved by using pulse-shape analysis with
segmented HPGe detectors will be sufficient to discriminate neutrons
from $\gamma$ rays.  The time-of-flight technique may be a
complementary procedure that can be used together with the tracking
based methods presented in this work.

\section{Monte Carlo simulations and \texorpdfstring{$\gamma$}-ray tracking}
\label{sec:liq}

The interaction points of neutrons and $\gamma$ rays in the HPGe
detectors of AGATA were simulated with the {\geant} toolkit versions
4.9.0 and 4.9.2 using the neutron cross section data libraries
G4NDL3.11 and G4NDL3.13, respectively. Within the low-energy models of
neutron scattering used by {\geant} prior to version 4.9.2, errors
related to the angular distribution of scattered neutrons as well as
errors in the energy distribution of the scattered neutrons, of the
recoiling nuclei, and of the $\gamma$ rays produced after the
inelastic scattering of neutrons, were reported on the {\geant}
problem tracking system \cite{Geant4-Bugzilla}. In version 4.9.2 of
{\geant} these bugs have been corrected. In our work we used for
{\geant} version 4.9.0 a modification of the {\geant} code that was
suggested by J. Ljungvall et al. \cite{2005NIMPA.546..553} and which
corrects for the errors in the energy distribution of the germanium
recoils. We noted that our results regarding inelastic scattering of 1
to 5 MeV neutrons on germanium obtained by {\geant} version 4.9.0,
with the correction suggested by J. Ljungvall et
al. \cite{2005NIMPA.546..553} were similar to the ones obtained with
{\geant} version 4.9.2.

The full AGATA array with 180 HPGe crystals was used in the {\geant}
simulations. The AGATA HPGe detectors were arranged in a $4\pi$
geometry around the central source position (= target position in an
in-beam experiment). In most of the simulations presented in this work
only the HPGe crystals (no other material) were included. In a few of
the simulations, the aluminium capsules in which the HPGe crystals are
mounted, were also included.

In a typical HIFE reaction the majority of the neutrons are emitted in
the energy range 1-5 MeV with a high-energy tail that extends up to
about 15 MeV. In this work we either simulated the emission of
mono-energetic neutrons or neutrons with a flat energy distribution
from 1 to 5 MeV, which covers the majority of the neutrons emitted
in a HIFE reaction.

The recoil energies of the Ge ions are not fully converted into
electron-hole pairs. This gives rise to the pulse-height defect (PHD)
\cite{Knoll} that was included in the simulations as suggested in
ref. \cite{2005NIMPA.546..553}.  The Ge ionization energy
$E_{\textrm{I}}$ is expressed as
\begin{equation} \label{eq:phd}
  E_{\mathrm{\textrm{I}}} = a \cdot E_{\mathrm{\textrm{R}}}^{\mathrm{b}}, 
  \;\; a = 0.21, \; \ b = 1.099, 
\end{equation}
where $E_{\textrm{R}}$ is the Ge recoil energy in the HPGe crystal.

The output of the {\geant} simulation program that contains the energy
and three-dimensional position of each of the interaction points in
AGATA was used as input to the {\mgt} $\gamma$-ray tracking program. At
this stage {\mgt} has no information whether the interaction is due to a
$\gamma$ ray or a neutron. Thus, the neutron interactions are treated
in the same way as the $\gamma$-ray interactions during the tracking
process.

Before applying the tracking with {\mgt} the interaction points that are
located closer to each other than 5~mm were packed together.  Their
energies were summed and their new position was calculated as the
energy-weighted average position. The packed interaction energies were
smeared by a Gaussian distribution with a full-width-at-half-maximum
(FWHM) corresponding to the typical intrinsic resolution of a large
volume HPGe detector.  The interaction positions were then
smeared by a Gaussian distribution having a FWHM of 5 mm at 100
keV. The smearing parameter had an energy dependence of (interaction point
energy)$^{-1/2}$. Interaction points that have an energy lower
than a threshold energy of 5 keV were removed. The selection of this
value is based on measured \cite{2009DINO} low-energy 
thresholds of the
actual AGATA HPGe detectors. For accepting an event in AGATA the deposited
energy in at least one crystal must be larger than about 30 to 50 keV. Such a trigger condition was tested and it was found that it had a negligible effect on the discrimination of events due to inelastic scattering of neutrons. Therefore, only the 5 keV energy threshold was used in this work.

The {\mgt} code uses a forward tracking method which is developed by
several groups in Europe\cite{Bazzacco-MGT,2004NIMPA.533..454} and in
the USA \cite{1999NIMPA.430..69}. The method is based on the
clusterization of interaction points belonging to the same initial
photon. The clusters contain interaction points with a limited angular
range, which is a result of the forward peaked Compton-scattering
cross section and the fast decrease of the mean free path in Ge as the
$\gamma$ ray looses its energy. An interaction is allowed to be a
member of several clusters since at the end of the tracking only the
cluster with the lowest figure-of-merit ($FM$) value is accepted. The
$FM$ calculation is based on a comparison of the energy of the
scattered $\gamma$ rays calculated by using the interaction energies
and by applying the Compton scattering formula to the positions of the
interaction points. The probabilities for Compton scattering, pair
production, and photo absorption (photoelectric effect), as well as
the probability for the $\gamma$ ray to travel the distances between
the interaction points in a cluster, are also included in the $FM$.

As an example, the equation for the $FM$ of a cluster with two
interaction points, the first one due to a Compton scattering and the
second one due to photo absorption, is given by
\begin{align} \label{eq:fm}
  FM = &
  \left(
  \frac{E_{\gamma,1}(\textrm{en}) - E_{\gamma,1}(\textrm{pos})}{E_{\gamma,0}}
  \right)^{2} \cdot \nonumber \\
  & \frac{d\sigma_{\textrm{KN}}/d\Omega}{P_{\textrm{Comp}}(E_{\gamma,0}) P(r_1)
    P_{\textrm{photo}}(E_{\gamma,1}) P(r_2)}.
\end{align}
Here $E_{\gamma,1}(\textrm{en})$ is the energy of the scattered
$\gamma$ ray after the first interaction point
\begin{equation} \label{eq:eg1}
  E_{\gamma,1}(\textrm{en}) = E_{\gamma,0} - E_1,
\end{equation}
with $E_{\gamma,0}$ being the sum of all interaction points (=
incident energy for a fully absorbed and correctly tracked $\gamma$
ray) and $E_1$ the energy deposited at the first interaction point.
$E_{\gamma,1}(\textrm{pos})$ is also the energy of the scattered
$\gamma$ ray after the first interaction point, but calculated by
using the Compton formula
\begin{equation} \label{eq:compt}
  E_{\gamma,1}(\textrm{pos}) = 
  E_{\gamma,0} \left[1 + \frac{E_{\gamma,0} (1 - \cos{\theta})}{m_{e} c^{2}}
  \right]^{-1},
\end{equation}
where the scattering angle of the $\gamma$ ray, $\theta$, is
calculated by using the position coordinates of the three interaction
points (target position, first and second interaction points in this
example) and $m_{e} c^{2}$ is the rest mass of the electron.
The probabilities for traveling the distances $r_1$ and $r_2$ are
given by
\begin{equation} \label{eq:p_lambda}
  P_{\lambda}(r_1)=\ex{-r_1/\lambda(E_{\gamma,0})},  \quad 
  P_{\lambda}(r_2)=\ex{-r_2/\lambda(E_{\gamma,1})},
\end{equation}
where $\lambda$ is the mean free path of $\gamma$ rays in Ge.  The
probabilities for Compton scattering and photo absorption are given by
\begin{align} \label{eq:p_comp_photo}
  P_{\textrm{Comp}} = 
  \frac{\sigma_{\textrm{Comp}}(E_{\gamma,0})}
       {\sigma_{\textrm{tot}}(E_{\gamma,0})},
       \nonumber \\
  P_{\textrm{photo}} = 
  \frac{\sigma_{\textrm{photo}}(E_{\gamma,1}(\textrm{en}))}
       {\sigma_{\textrm{tot}}(E_{\gamma,1}(\textrm{en}))}.
\end{align} 
The differential cross section $d{\sigma_{\textrm{KN}}}/d{\Omega}$,
which does not have an important effect on our tracking results, is
the Klein-Nishina Compton scattering cross section for unpolarized
radiation.  

During the tracking process the $FM$ values of all possible
permutations of interaction points are evaluated. The one with the
lowest $FM$ value is chosen as an accepted tracked $\gamma$ ray, but
only of its $FM$ value is $<1$. It should be noted that similar but
not identical $FM$ definitions as in eq. \ref{eq:fm} are given in
refs.  \cite{2004NIMPA.533..454,1999NIMPA.430..69}. These three $FM$
equations are implemented as options in {\mgt} and they give similar
results \cite{2009DINO}.

\begin{table*}[htb!]
  \renewcommand{\arraystretch}{1.2}
  \centering
  \caption{Energies and multiplicities of $\gamma$ rays and neutrons
    in the simulated data sets used in this work. The Ge isotope
    enrichment and the used pulse-height defect (PHD) correction are
    also given.  The $\gamma$-ray cascade is a ``rotational band''
    with 10 transitions from $E_{\gamma} = 150$ keV to 1450 keV with
    $\Delta E_{\gamma} = 150$ keV.}
  \label{tab:01}
  \begin{tabular*}{\textwidth}{@{\extracolsep{\fill}}ccccccc}
    \hline
    \multirow{2}{*}{Data set number} &  
    \multirow{2}{*}{Ge enrichment} &
    \multicolumn{2}{c}{$\gamma$-rays}  &
    \multicolumn{2}{c}{neutrons} & 
    \multirow{2}{*}{PHD} \\
    \cline{3-6}
    & & $E_{\gamma}$ [keV] & $M_{\gamma}$ &
    $E_{\textrm{n}}$ [MeV]  & $M_{\textrm{n}}$ & \\
    \hline
    1  & $^{74}$Ge          & 596     & 1  & - & 0 & - \\
    2  & $^{\textrm{nat}}$Ge & 1000    & 1  & - & 0 & -  \\
    3  & $^{\textrm{nat}}$Ge & cascade & 10 & - & 0 & -   \\
   \hline
   4  & $^{74}$Ge          & - & 0 & 1    & 1  &  no \\
   5  & $^{74}$Ge          & - & 0 & 1, 3, 5    & 1 & yes   \\
   6  & $^{\textrm{nat}}$Ge  & - & 0 & 1-5  & 1  & yes \\
   7  & $^{\textrm{nat}}$Ge  & cascade & 10  &  1-5 & 6 & yes  \\
   \hline
  \end{tabular*}
\end{table*}

The properties of the data sets used in this work are summarized in
table \ref{tab:01}. Data sets 1, 4, and 5 were produced in order to
develop the methods for the discrimination of neutron and $\gamma$-ray
interaction points. Data set 6 contains events produced with a flat
distribution of neutron energies from 1 to 5 MeV, which well enough
emulates the energy distribution of neutrons emitted in a typical HIFE
reaction. The results obtained with data set 6 were compared to the
ones obtained with data set 2, which contains events generated by 1
MeV $\gamma$ rays emitted from the center of AGATA. The energy 1 MeV
was chosen to represent a typical average $\gamma$-ray energy.
Finally, a more realistic simulation was carried out with a flat
distribution of neutron energies between 1 and 5 MeV emitted from the
center of AGATA together with a $\gamma$-ray cascade representing a
``rotational band'' populated after a HIFE reaction, data set 7.  The
results obtained with data set 7 were compared to the results of data
set 3, which does not contain any neutron emissions.

\section{Ge recoil energy distributions} \label{ss:rec} 

The n + $^{\textrm{nat}}$Ge reaction cross section is more or less
constant for neutron energies in the range 1-5 MeV and it is dominated
by elastic scattering \cite{2005NIMPA.546..553}.  Inelastic scattering
has the second largest cross section, being a factor of 8 and 1.3
smaller than elastic scattering at the neutron energies 1 MeV and 5
MeV, respectively. In this energy interval the cross sections of other
reactions are much smaller and have in the context of this work been
neglected. In the elastic scattering process, some of the neutron
energy is deposited in the detector via the recoiling Ge nuclei. These
energies give rise to single-hit clusters during the $\gamma$-ray
tracking process and they are mostly eliminated from the data since
their $FM$ value is not good enough and they do not qualify as
photo-absorption points. Inelastic scattering of neutrons is more
complicated since in this case the recoiling Ge nuclei usually also
emit $\gamma$ rays.  These $\gamma$ rays cause an unwanted background
in the $\gamma$-ray spectra.

\begin{figure}[htb!] 
  \centering
  \includegraphics[width=0.55\columnwidth,angle=90]{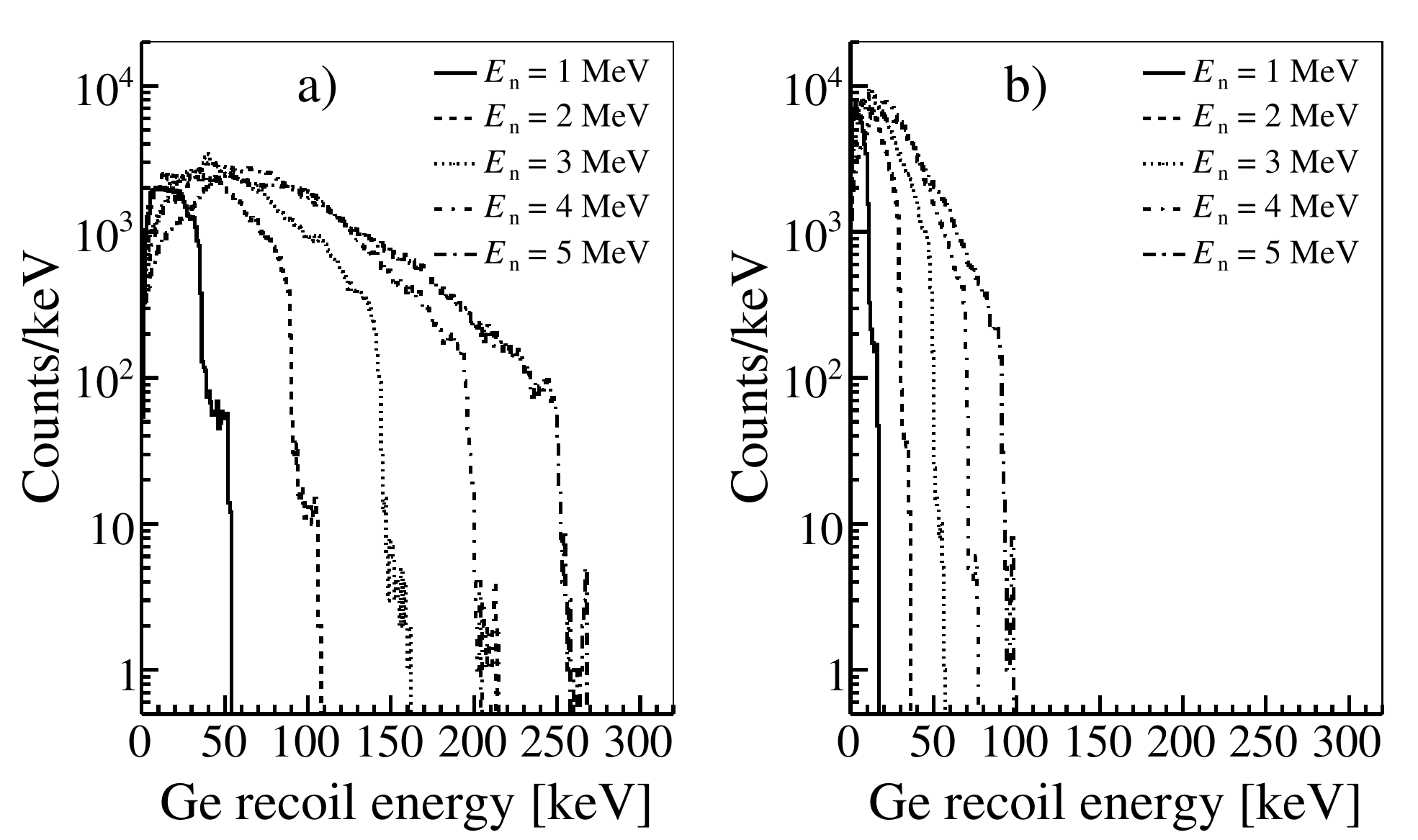}
  \caption{Recoil energies of $^{\textrm{nat}}$Ge after inelastic
    scattering of neutrons with energies from 1 to 5 MeV, a) without
    and b) with PHD correction.}
  \label{fig:01}
\end{figure}

Simulations of the Ge recoil energy distributions were performed by
using the most recent version of {\geant} (version
4.9.2). Distributions for the different Ge isotopes as well as for
$^{\textrm{nat}}$Ge were investigated. The shapes of the recoil energy
distributions are determined by the fairly isotropic angular
distributions and by the reaction kinematics. For scattering of 1 MeV
neutrons on $^{\textrm{nat}}$Ge, the recoil energy extends from 0 to
57 keV for elastic scattering.  The recoil energy distributions due to
inelastic scattering of 1 to 5 MeV neutrons on $^{\textrm{nat}}$Ge are
shown in fig. \ref{fig:01}a. The recoil energies due to inelastic
scattering of 1 MeV neutrons extends from 0 to about 35 keV for the
even-even isotopes $^{70}$Ge, $^{72}$Ge, $^{74}$Ge, and $^{76}$Ge and
to 53 keV for $^{73}$Ge.

Figure \ref{fig:01}b shows the effect of the PHD correction
(eq. \ref{eq:phd}) on the recoil energies.  The maximum recoil energy
after inelastic neutron scattering with $E_{\textrm{n}} = 1 $ MeV is
only 16.7 keV after the PHD correction.  With a low-energy threshold
of 5 keV, 52 \% of the neutron interaction points for 1 MeV neutrons
and 5 \% for 5 MeV neutrons will not be detected after the PHD
correction.

\section{Number of interaction points} \label{s:npt} 

\begin{figure}[htb!] 
  \centering
  \includegraphics[width=0.95\columnwidth]{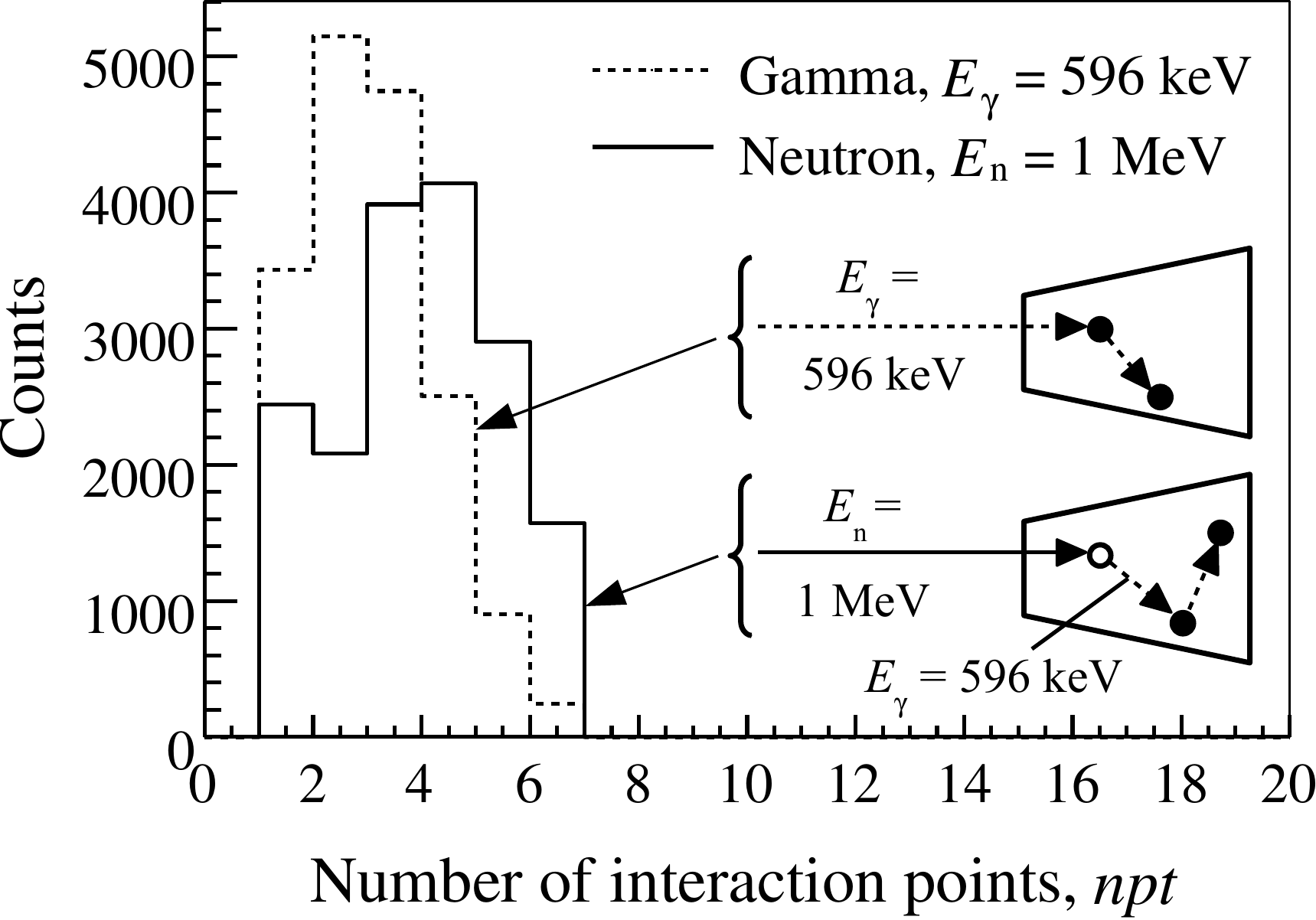}
  \caption{Number of interaction points in $^{74}$Ge after
    tracking. The spectra
    are normalized to have the same number of counts.}
  \label{fig:02}
\end{figure}

In a simplified example, we have investigated whether the $\gamma$-ray
tracking code can discriminate between two types of $\gamma$-rays with
the same energies, namely 596 keV.  The first type of $\gamma$ ray is
emitted from the center of AGATA (data set 1), while the second type
is created within a HPGe detector after inelastic scattering of 1 MeV
neutrons on $^{74}$Ge (data set 4). This is illustrated in the inset
of fig. \ref{fig:02}. To avoid creating $\gamma$ rays of other
energies than 596 keV from the inelastic neutron scattering, the AGATA
HPGe detectors were in this simulation configured to consists only of
$^{74}$Ge isotopes. With 1 MeV neutrons one can only populate the
first excited $2_1^+$ state at 596 keV in $^{74}$Ge (the second
excited state is at 1204 keV), which then de-excites to the $0_1^+$
ground state by emitting a 596 keV $\gamma$ ray.

The number of interaction points ($npt$) in $^{74}$Ge determined by
the tracking algorithm is shown in fig. \ref{fig:02}. When the
incoming particle is a 1 MeV neutron $npt$ is increased on the average
by one unit compared to the events where the incoming particle is a
596 keV $\gamma$ ray. This increase in the $npt$ value is due to the
neutron interaction point that is included in the {\mgt} cluster
together with the $\gamma$-ray interaction points.

\begin{figure}[htb!] 
  \centering
  \includegraphics[width=0.95\columnwidth]{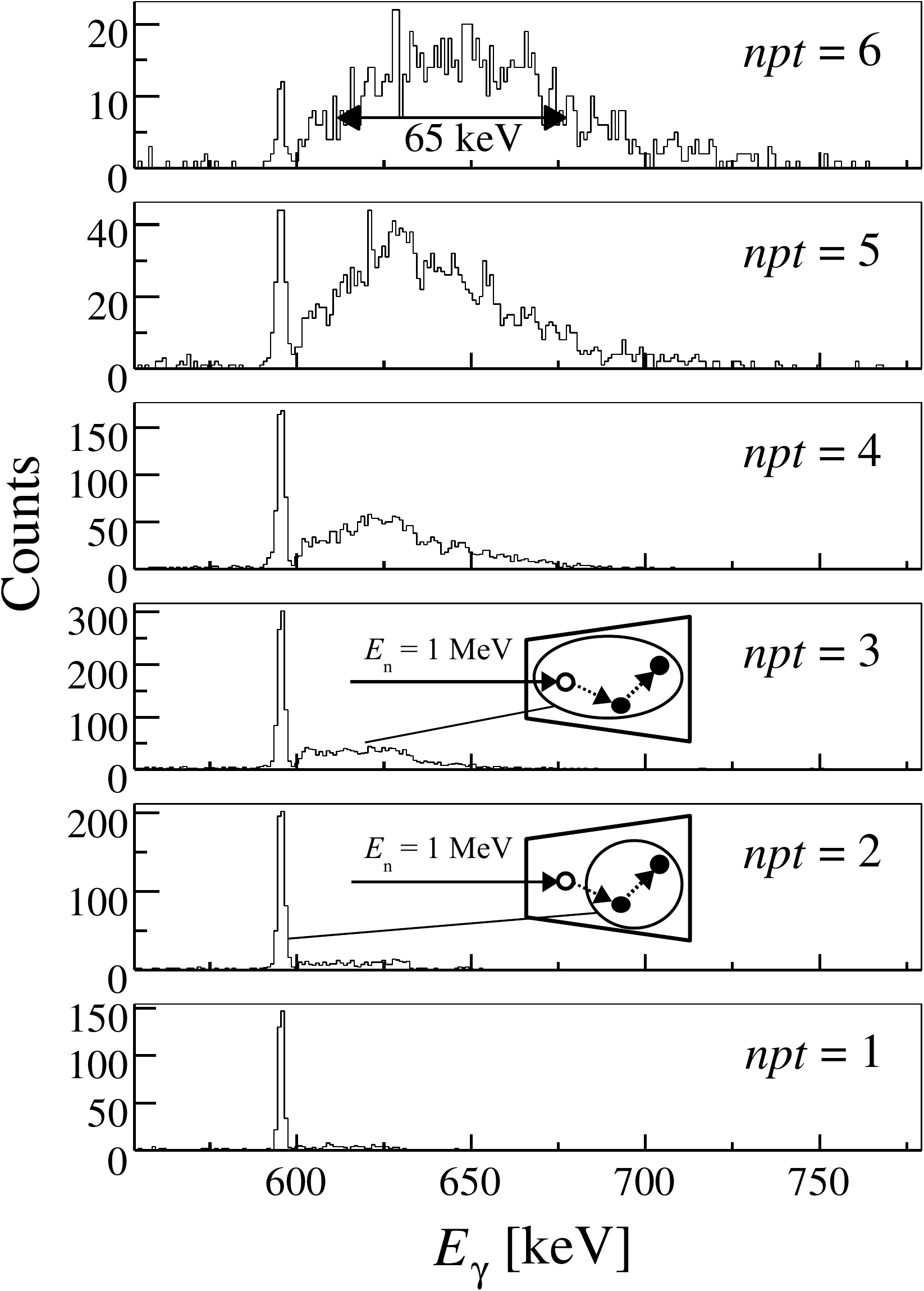}
  \caption{Gamma-ray spectra due to inelastic scattering of 1 MeV
    neutrons on $^{74}$Ge (data set 4) obtained after $\gamma$-ray
    tracking.  In sub-figures $npt=2$ and $npt=3$, an example with
    three interactions points is illustrated. The first interaction
    point is due to the recoiling Ge nucleus after inelastic
    scattering of the neutron (open circle). The second and third
    interaction points (filled circles) are due to Compton scattering
    and photo absorption, respectively, of the $\gamma$ ray produced
    in the decay of the excited Ge nucleus.  If the neutron
    interaction point is not measured or not included in the
    {\mgt} cluster, the 596 keV peak in the $npt = 2$ spectrum
    is incremented. Otherwise, if the recoil energy is measured and
    included in the cluster, the bump in the $npt = 3$ spectrum is
    incremented.}
  \label{fig:03}
\end{figure}

Simulated and tracked $\gamma$-ray spectra obtained after inelastic
scattering of 1 MeV neutrons on $^{74}$Ge are shown for different
$npt$ values in fig. \ref{fig:03}.  Apart from the peak at 596 keV, a
bump is visible at higher $\gamma$-ray energies. This bump originates
from the {\mgt} clusters where the interaction point due to the
Ge recoil is included in the cluster and its energy is added to 596
keV. As $npt$ increases, it becomes more likely that one of the
interaction points is due to the scattering of neutrons and the bump
becomes more prominent. At $E_{\textrm{n}} = 1$ MeV the width of the
recoil energy distribution is 35 keV (see section \ref{ss:rec}). The
FWHM of the bump in fig. \ref{fig:03} is about 65 keV for $npt = 6$,
which indicates that there may be two or even three neutron
interaction points in the clusters, apart from the $\gamma$-ray
interaction points.

\section{Neutron-\texorpdfstring{$\gamma$} discrimination methods} \label{ss:methods} 

Based on a comparison of data sets 1 and 4, three methods to
distinguish between two types of $\gamma$ rays, were investigated.

\subsection{Method 1: Energy deposited in the first and second
  interaction points} \label{sss:method1}

\begin{figure}[htb!]
  \centering
      \includegraphics[width=0.65\columnwidth,angle=90]{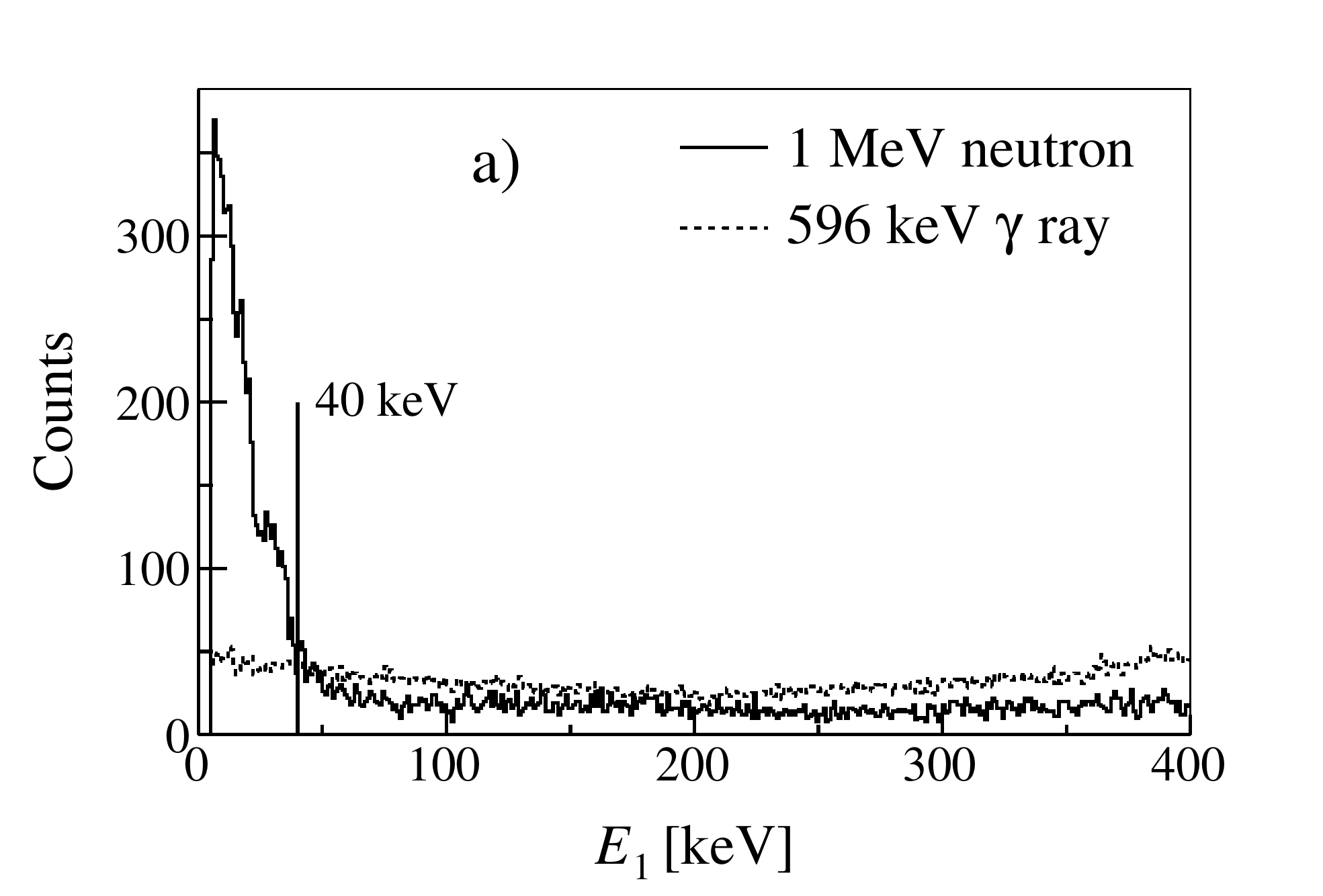}
      \includegraphics[width=0.65\columnwidth,angle=90]{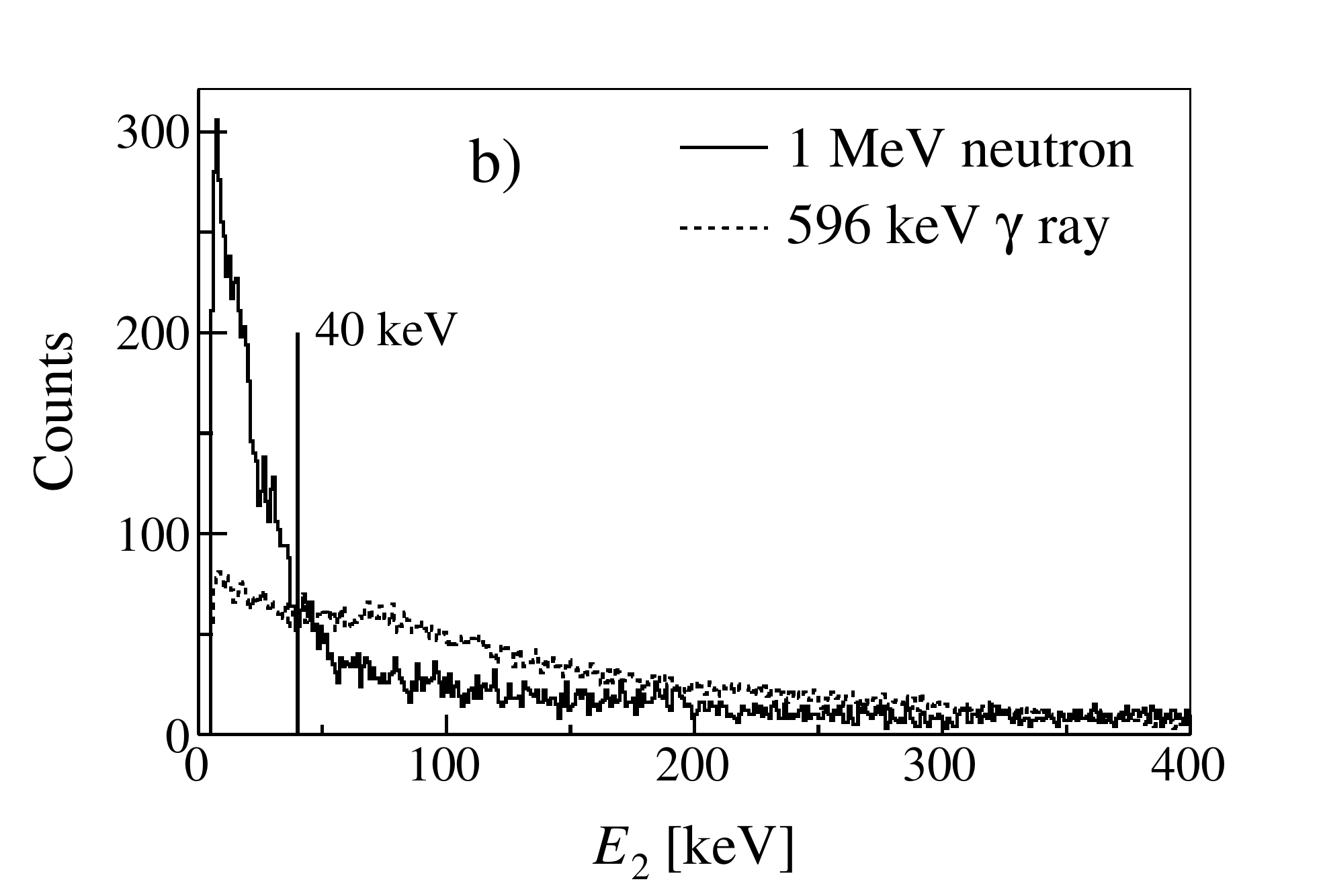} 
  \caption{Energy deposited in a) the first and b) the second
    interaction points. The spectra were obtained by sending 596 keV
    $\gamma$ rays (dashed lines, data set 1) and 1 MeV neutrons (solid
    lines, data set 4) from the center of AGATA. The HPGe detectors
    were made of $^{74}$Ge isotopes in these simulations. The spectra
    in a) and b) have been normalized to have the same number of
    counts.}
  \label{fig:04}
\end{figure}

The first method is based on the low energy deposition in the
detectors by the recoiling Ge nuclei after inelastic scattering of
neutrons. Distributions of the energies of the first ($E_1$) and
second interaction points ($E_2$) for each {\mgt} cluster are
shown in fig. \ref{fig:04}. For 1 MeV neutrons both the $E_1$ and
$E_2$ spectra have large abundances of counts at low energies, below
about 40 keV, compared to the spectra of 596 keV $\gamma$ rays emitted
from the center of AGATA. With a condition (gate) of $E_1$ and $E_2$
to be $>40$ keV one can discriminate a large number of the cases in
which the incoming particle is a neutron compared to a $\gamma$ ray.
For the second interaction point the shape of the spectrum for
incoming 596 keV $\gamma$ rays is somewhat different than for the
first interaction point, in particular at low energies.  With a gate
at $E_2 > 40$ keV slightly more of the photopeak efficiency of the 596
keV peak is lost as compared to the same gate on $E_1$. In the
simulations presented below the gate on $E_2$ was therefore set at a
smaller value of the interaction point energy. The energies of the
third interaction point were also studied but they were not used for
gating in this work since they caused an even bigger loss in the
photopeak efficiency than $E_2$.

\subsection{Method 2: Difference in incoming direction of the 
  \texorpdfstring{$\gamma$} ray} \label{sss:method2}

\begin{figure}[htb!] 
  \centering
  \includegraphics[width=0.95\columnwidth]{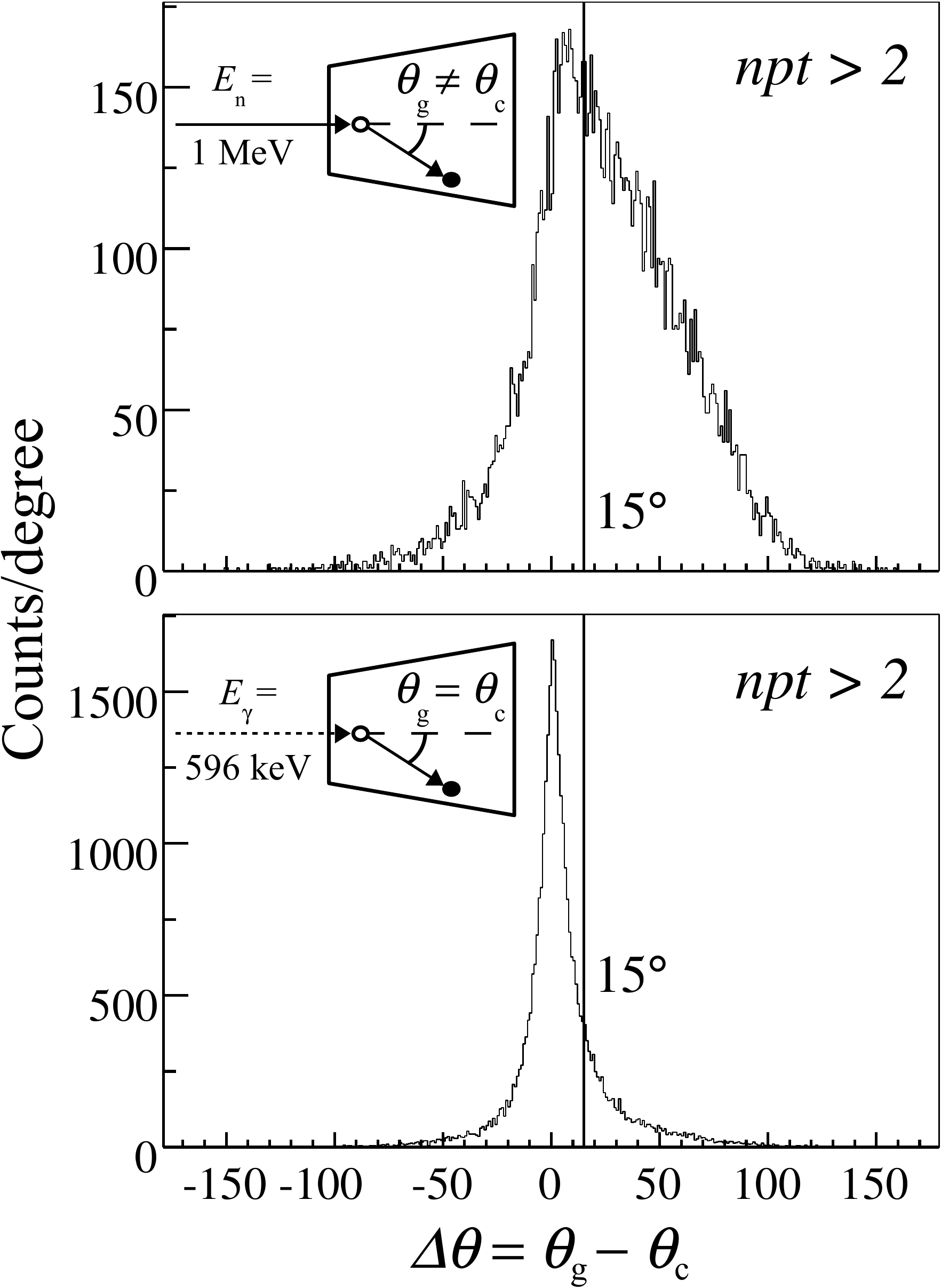}
  \caption{Distribution of the difference $\Delta\theta =
    \theta_{\textrm{g}} - \theta_{\textrm{c}}$ obtained when emitting
    596 keV $\gamma$ rays (lower, data set 1) and 1 MeV neutrons
    (upper, data set 4) from the center of AGATA. The HPGe detectors
    were made of $^{74}$Ge isotopes in these simulations. The angle
    $\theta_{\textrm{g}}$ is obtained from the position of the
    interaction points whereas $\theta_{\textrm{c}}$ is obtained from
    the Compton scattering formula.}
  \label{fig:05}
\end{figure}

The second method is based on the difference in the incoming direction
of the initial $\gamma$ rays. If they come from random positions in
space, they must be produced by inelastic neutron scattering in the
HPGe detectors. If they come from the center of the array, they are
likely not produced by reactions in the detectors. Two different
angles, $\theta_{\textrm{g}}$ and $\theta_{\textrm{c}}$, are extracted
and compared in this investigation in a similar way that was used in
ref. \cite{2005NIMPA.550..379}. As illustrated in fig. \ref{fig:05},
$\theta_{\textrm{g}}$ is the geometric angle between the line that
goes from the center of the detector array to the first interaction
point and the line that connects the first interaction point to the
second one. Thus, $\theta_{\textrm{g}}$ is determined from the
positions of the interaction points relative to the center of
AGATA. The angle $\theta_{\textrm{c}}$ is the scattering angle
calculated from the Compton scattering formula using the total energy
deposited by the $\gamma$ ray (sum of all energies in the cluster) and
the energy of the first interaction point obtained after $\gamma$-ray
tracking.

If the $\gamma$ ray comes from the center of AGATA we have
$\theta_{\textrm{g}} = \theta_{\textrm{c}}$ and we expect to see a
sharp distribution of the difference $\Delta\theta =
\theta_{\textrm{g}} - \theta_{\textrm{c}}$ centered at
$0^{\circ}$. The width of this distributions is due to the errors in
the determination of the energy and position of the interaction
points. Such a distribution is seen in the lower part of
fig. \ref{fig:05}. The upper part of the same figure shows the
distribution of $\gamma$ rays produced by inelastic scattering of 1
MeV neutrons on $^{74}$Ge. This distribution is wider than the first
one and it is asymmetric around $0^{\circ}$.  The asymmetry is due to
the low-energy interaction points (due to Ge recoils), which are
accepted by the tracking code as forward Compton scattered $\gamma$
rays with a small $\theta_{\textrm{c}}$ angle giving an excess of
positive $\Delta\theta$ values.  In this work we have used the
condition $\Delta\theta > 15^{\circ}$, in order to discriminate
neutrons and $\gamma$ rays. If $\Delta\theta$ is larger than
$15^{\circ}$ the cluster is assumed to contain a $\gamma$ ray from
inelastic neutron scattering.  It should be mentioned that the neutron
rejection obtained by this method can be improved with a better
interaction position resolution. An increased position resolution
leads to a smaller width of the distribution in the lower part of
fig. \ref{fig:05}, which allows the use of a lower value of the gate
on $\Delta\theta$.

\subsection{Method 3: Selection based on the \texorpdfstring{$FM$} value} 
\label{sss:method3}

The third method is based on a comparison of figure-of-merit values
($FM$) of the clusters determined by the {\mgt} tracking
algorithm when the initial particles are 1 MeV neutrons and 596 keV
$\gamma$ rays emitted from the center of AGATA. In the {\mgt}
program a cluster is accepted only if all its interactions are valid
as one of the three $\gamma$-ray processes: Compton scattering, photo
absorption, or pair production.  The validity depends on the energies
deposited at each interaction point, the probability of the process to
happen and the interaction length before the process takes place.  A
cluster that contains a neutron interaction point may be accepted ($FM
< 1$) only if the neutron interaction point accidentally qualifies as
a $\gamma$-ray interaction point. Such a cluster is, however, expected
to give a higher $FM$ value (eq. \ref{eq:fm}) as compared to a cluster
with only $\gamma$-ray interactions.

\begin{figure}[htb!] 
  \centering
  \begin{center}
    \includegraphics[width=0.75\columnwidth,angle=90]{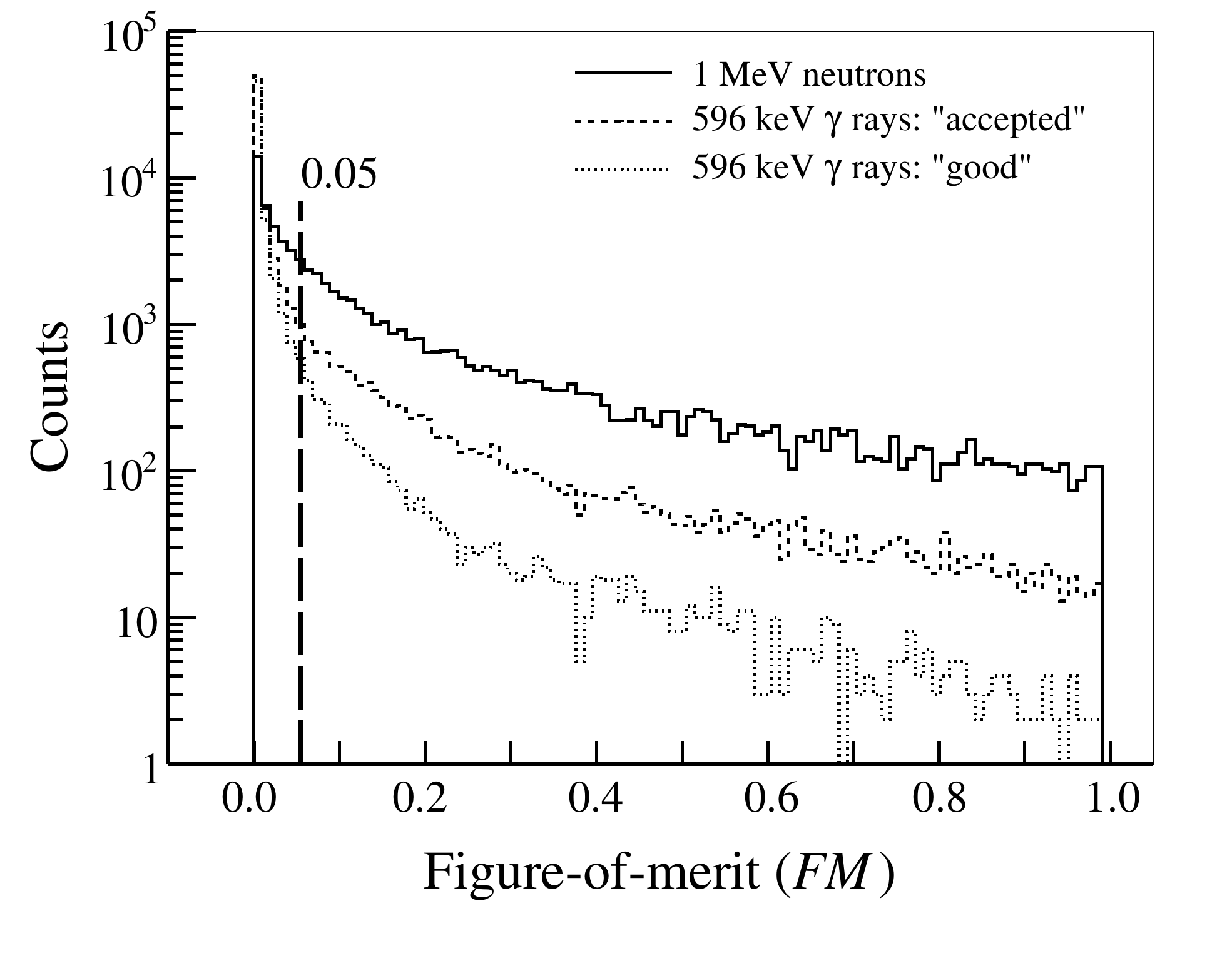}
  \end{center}
  \caption{Figure-of-merit ($FM$) values of the {\mgt} code when the
    initial particles are 596 keV $\gamma$ rays (dashed lines, data
    set 1) and 1 MeV neutrons (solid line, data set 4) emitted from
    the center of AGATA. The HPGe detectors were made of $^{74}$Ge
    isotopes in these simulation. The label ``accepted'' $\gamma$ rays
    indicates the $\gamma$ rays which fulfill the requirement $FM <
    1$, whereas the ``good'' $\gamma$ rays in addition have an
    energy of 596 keV. The histogram of the 1 MeV neutrons and the
    ``accepted'' 596 keV $\gamma$ rays have been normalized to have
    the same number of counts.  }
  \label{fig:06}
\end{figure}

Figure \ref{fig:06} shows a comparison of the $FM$ values of the
clusters constructed by the {\mgt} code after the emission of
596 keV $\gamma$ rays from the center of AGATA and the $FM$ values of
the clusters constructed after the emission of 1 MeV neutrons. In this
figure both ``accepted'' and ``good'' $\gamma$ rays belong to clusters
with $FM < 1$. The energy of the ``good'' $\gamma$ rays, however, has
the value of 596 keV, i.e. it is equal to the incoming $\gamma$-ray
energy.  The $FM$ distribution of the ``good'' $\gamma$ rays is
strongly peaked at small $FM$ values and it decreases much faster than
the distribution due to 1 MeV neutrons.  With the help of
fig. \ref{fig:06} we decided to use the gate $0.05 < FM < 1$ in order
to discriminate between neutrons and $\gamma$ rays.

\section{Comparison of interaction distances} \label{ss:int_dist}

\begin{figure}[htb!] 
  \centering
  \begin{center}
    \includegraphics[width=0.75\columnwidth,angle=90]{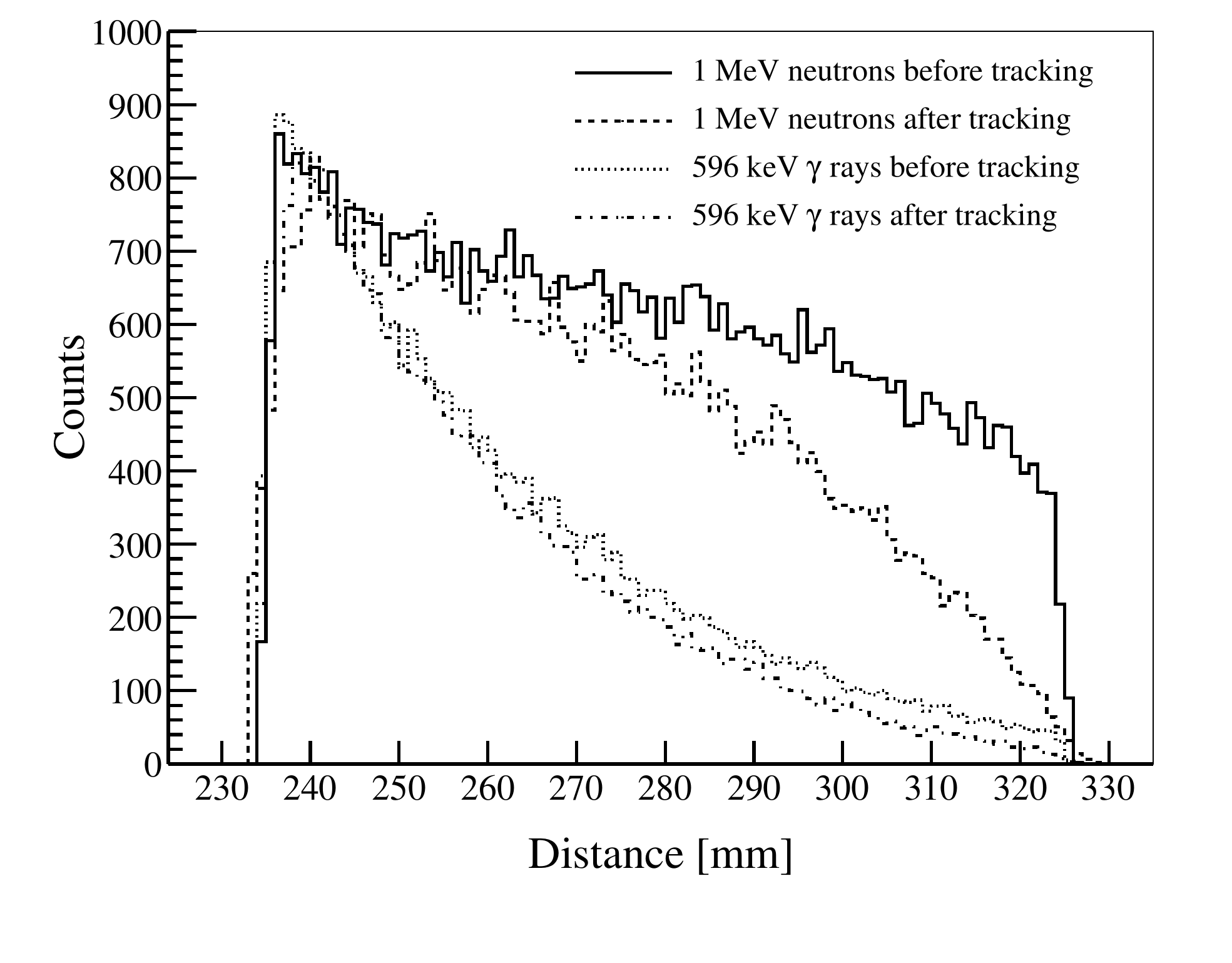}
  \end{center}
  \caption{Interaction distances measured from the center of AGATA to
    the first interaction point for 596 keV $\gamma$ rays (data set 1)
    and for inelastic scattering of 1 MeV neutrons (data set 4) on
    $^{74}$Ge.  The histograms have been arbitrarily normalized.  }
  \label{fig:07}
\end{figure}

We have also investigated and compared the distances from the center
of AGATA to the first interaction point for 596 keV $\gamma$ rays
emitted a) from the center of AGATA (data set 1) and b) from inelastic
scattering of 1 MeV neutrons (data set 4) on $^{74}$Ge. The comparison
was done both for the real first interaction point as obtained from
the {\geant} simulation (before tracking) and for the first
interaction point as obtained by {\mgt} (after tracking). The results
are shown in fig. \ref{fig:07}. For 596 keV $\gamma$ rays emitted from
the center of AGATA, the distributions of the distances to the first
interaction point before and after tracking are very similar. The
centroid of the distribution is at 26 cm, which corresponds to the
mean free path of 596 keV $\gamma$ rays in Ge, 2.5 cm, plus the 23.5
cm distance from the center to the front face of the Ge detectors.
For 1 MeV neutrons there is an abundance of events for which the real
first interaction distances (before tracking) are well above 26 cm
(see fig. \ref{fig:07}). However, the events with too large
interaction distances are eliminated by {\mgt} because the
$P_{\lambda}(r_1)$ value becomes too small, which leads to an $FM$
value larger than 1 (see eqs.  \ref{eq:fm}-\ref{eq:p_lambda}).

We have found that setting different conditions on the interaction
distances in {\mgt} rejects rather few neutron events and leads to a
relatively large loss of photopeak efficiency, compared to what we
have achieved with the other three methods mentioned above.  Since
this is also true for higher neutron energies (up to 5 MeV), it is
concluded here that gating on the interaction distances is not
beneficial for improving the neutron-$\gamma$ discrimination.
 
\section{Applications of the neutron-\texorpdfstring{$\gamma$} 
  discrimination methods} \label{sec:discr}

The results of the applications of the different neutron-$\gamma$
discrimination methods, as described in section \ref{ss:methods}, are
presented in this section.

\subsection{596 keV \texorpdfstring{$\gamma$} rays and 1 MeV neutrons
  (data set 1 and 4)} \label{ss:ds14}

\begin{figure}[htb!] 
  \centering
  \begin{center}
    \includegraphics[width=0.8\columnwidth,angle=90]{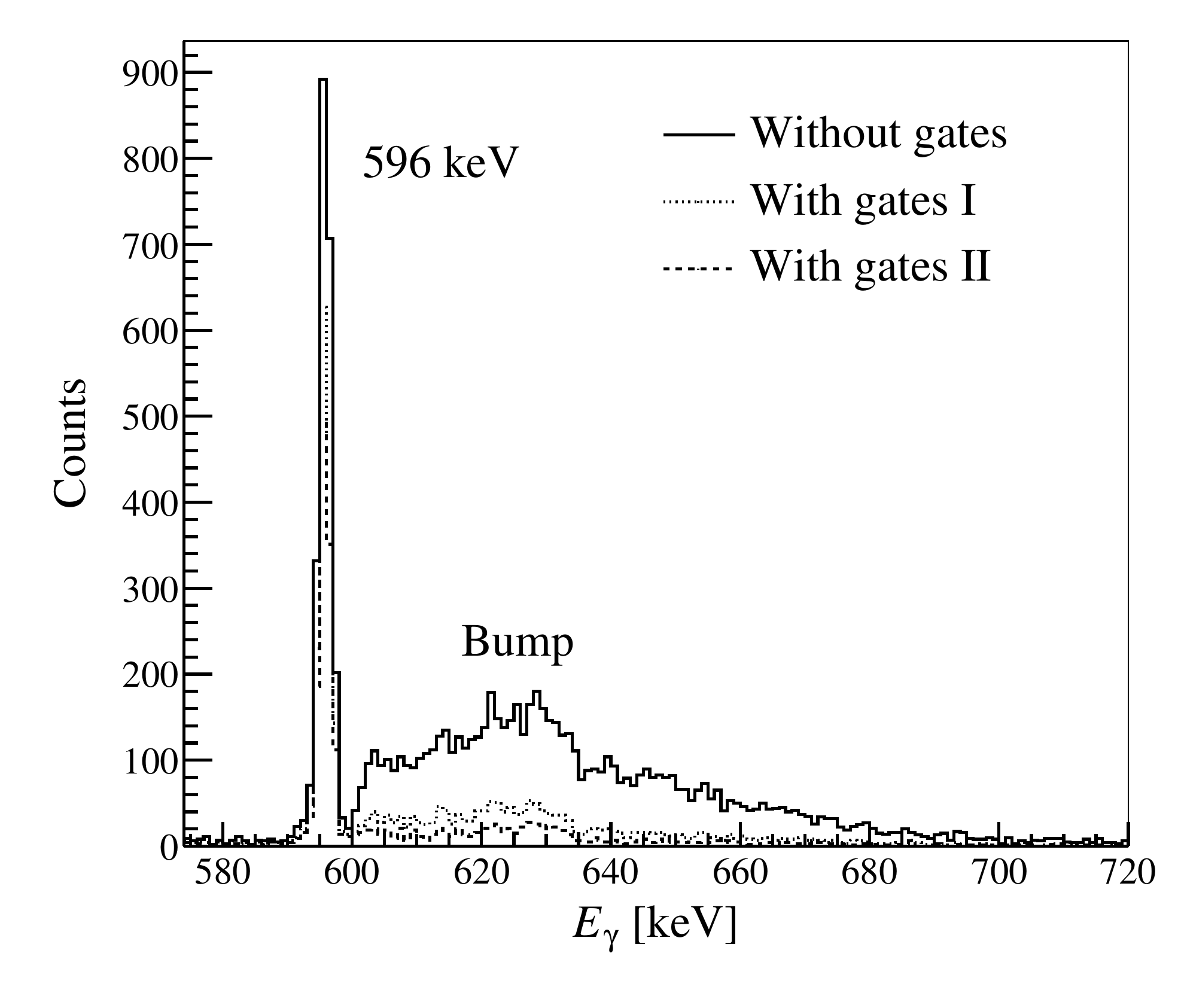}
  \end{center}
  \caption{Reduction of the 596 keV peak and its bump after applying
    different neutron-$\gamma$ discrimination methods. The spectra
    were obtained by sending 1 MeV neutrons on $^{74}$Ge (data set 4).
    See caption of table \ref{tab:02} for the definition of the gates
    I and II.  }
  \label{fig:08}
\end{figure}

Fig. \ref{fig:08} shows tracked $\gamma$-ray spectra in the region
around the 596 keV peak and its bump obtained from events generated by
inelastic scattering of 1 MeV neutrons on $^{74}$Ge (data set 4). The
dashed and dotted lines indicate what is left after applying different
gates. Quantitative values of the achieved reductions are shown in
table \ref{tab:02} together with the reduction of the photopeak
efficiency of the 596 keV peak for 596 keV $\gamma$ rays emitted from
the center of AGATA (data set 1). The aim of the gating is to reduce
the 596 keV peak and its bump in the neutron induced spectra as much
as possible while keeping the photopeak efficiency of the 596 keV peak
in the $\gamma$-ray induced spectra as large as possible.
\begin{table*}[htb!]
  \renewcommand{\arraystretch}{1.2}
  \caption{Reduction of counts in percent of the 596 keV peak, its
    bump, and of the total spectrum after applying different
    neutron-$\gamma$ discrimination gates.  The results were obtained
    by sending 1 MeV neutrons (data set 4) and 596 keV $\gamma$ rays
    (data set 1) on $^{74}$Ge.  The gates used were $E_1< 40$ keV,
    $E_2 < 30$ keV, $\Delta\theta > 15^{\circ}$, $0.05 < FM < 1$. The
    gate combinations ($E_1$, $FM$) and ($E_1$, $E_2$, $\Delta\theta$,
    $FM$) are labelled I and II, respectively, and were created by
    using a logical or of the individual gates.  }
  \label{tab:02}
  \begin{tabular*}{\textwidth}{@{\extracolsep{\fill}}lcccc}
    \hline
    \multirow{3}{*}{Gates} 
    & \multicolumn{3}{c}{1 MeV neutrons (data set 4)}
    & 596 keV $\gamma$ rays (data set 1)\\
    \cline{2-5} & 596 keV peak & Bump & Total & Peak \\
    \hline
    $E_1$
    & 6 & 63 & 40 & 6 \\
    $\Delta\theta$
    & 26 & 65 & 46 & 11 \\
    $FM$
    & 28 & 55 & 44 & 5 \\
    I: ($E_1$, $FM$)
    & 32 & 77 & 58 & 11 \\
    II: ($E_1$, $E_2$, $\Delta\theta$, $FM$)
    & 47 & 92 & 69 & 22 \\
    \hline
  \end{tabular*}
\end{table*}

By using gate set II ($E_1$, $E_2$, $\Delta\theta$, $FM$) the total
spectrum is reduced by 69 \%, the 596 keV transition by 47 \% and the
bump by 92 \% in the 1 MeV neutron induced spectrum.  This causes a
loss of 22 \% in the 596 keV full-energy peak of the spectrum obtained
by sending 596 keV $\gamma$ rays form the center of AGATA.  The two
neutron-$\gamma$ discrimination methods that contribute least to the
reduction of the photopeak efficiency of $\gamma$ rays emitted from
the center of AGATA are the $E_1$ and $FM$ tests. If we use an or of
gates based on these two methods (gate set I) the total neutron
induced spectrum is reduced by 58 \% while the peak is reduced by 32
\% and the bump by 77 \%. This causes a loss of 11 \% of the 596 keV
full-energy peak in the $\gamma$-ray induced spectrum.

Note that the three methods for neutron-$\gamma$ discrimination do not
exclude each other and there is an overlap of neutron events rejected
by the different gates. For example, the application of gate set I
gives a total neutron rejection of 58 \%, which is lower than the sum
of the results given by the $E_1$ and $FM$ gates separately. It is a
different situation when the same gate set is applied on the 596 keV
$\gamma$ rays emitted from the center of AGATA. In this case there is
no overlap of the events rejected by the different methods and the
total loss in the full-energy peak is a sum of the losses obtained by
the $E_1$ and $FM$ gates separately, as is seen in the last column of
table \ref{tab:02}.

\subsection{Pulse-height defect correction} \label{ss:phd}

The energies of the neutron interaction points are reduced
considerably when the PHD correction is included in the simulations,
see section \ref{sec:liq}. If the interaction energy is below the
low-energy threshold of 5~keV it will not be measured and the
rejection methods, in particular method 1 which uses $E_1$ (and
$E_2$), will not give as good results as the ones shown in table
\ref{tab:02}.

\begin{figure}[htb!] 
  \centering
  \begin{center}
    \includegraphics[width=0.47\columnwidth,angle=90]{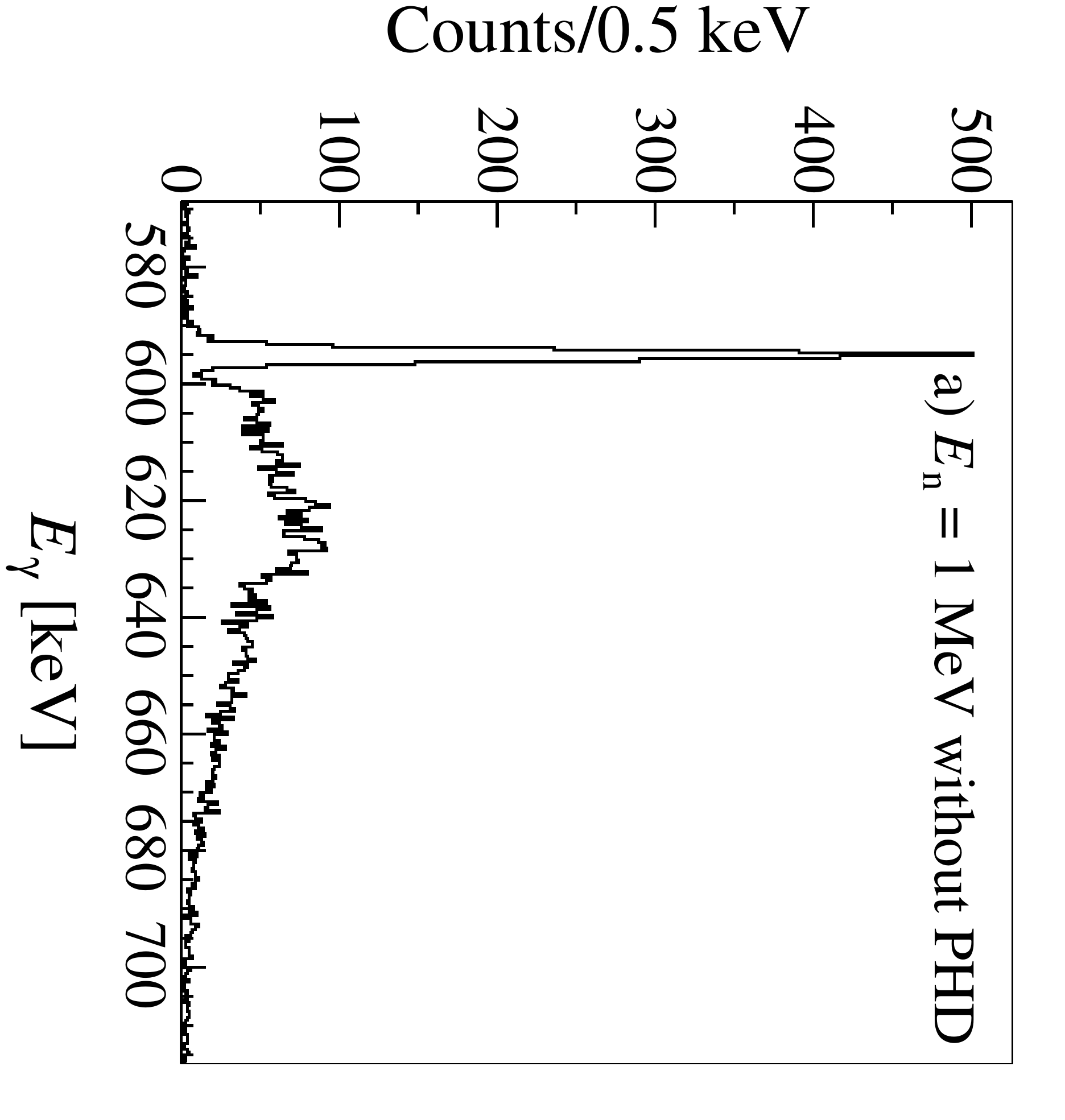}
    \includegraphics[width=0.47\columnwidth,angle=90]{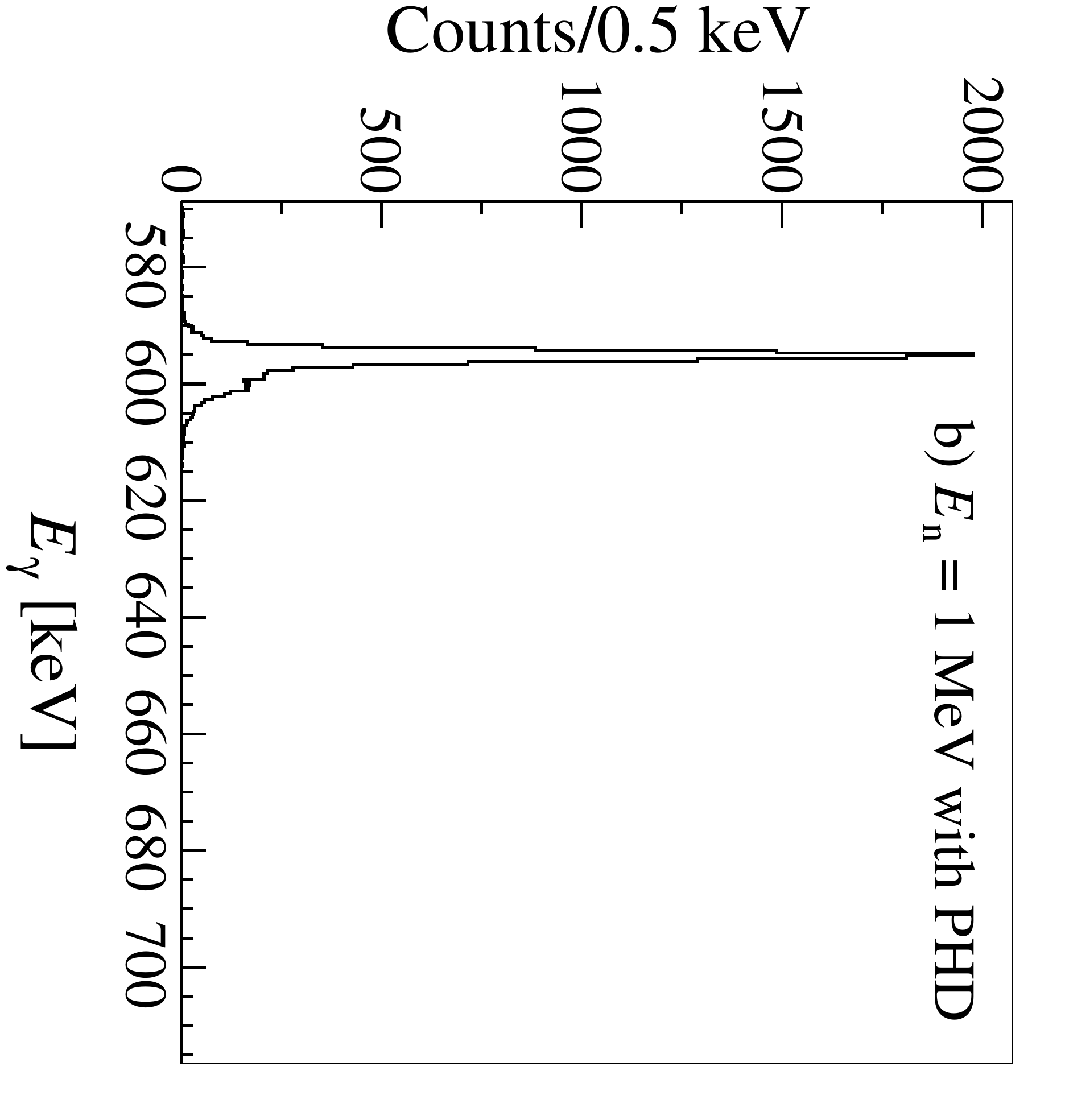} \\
    \includegraphics[width=0.47\columnwidth,angle=90]{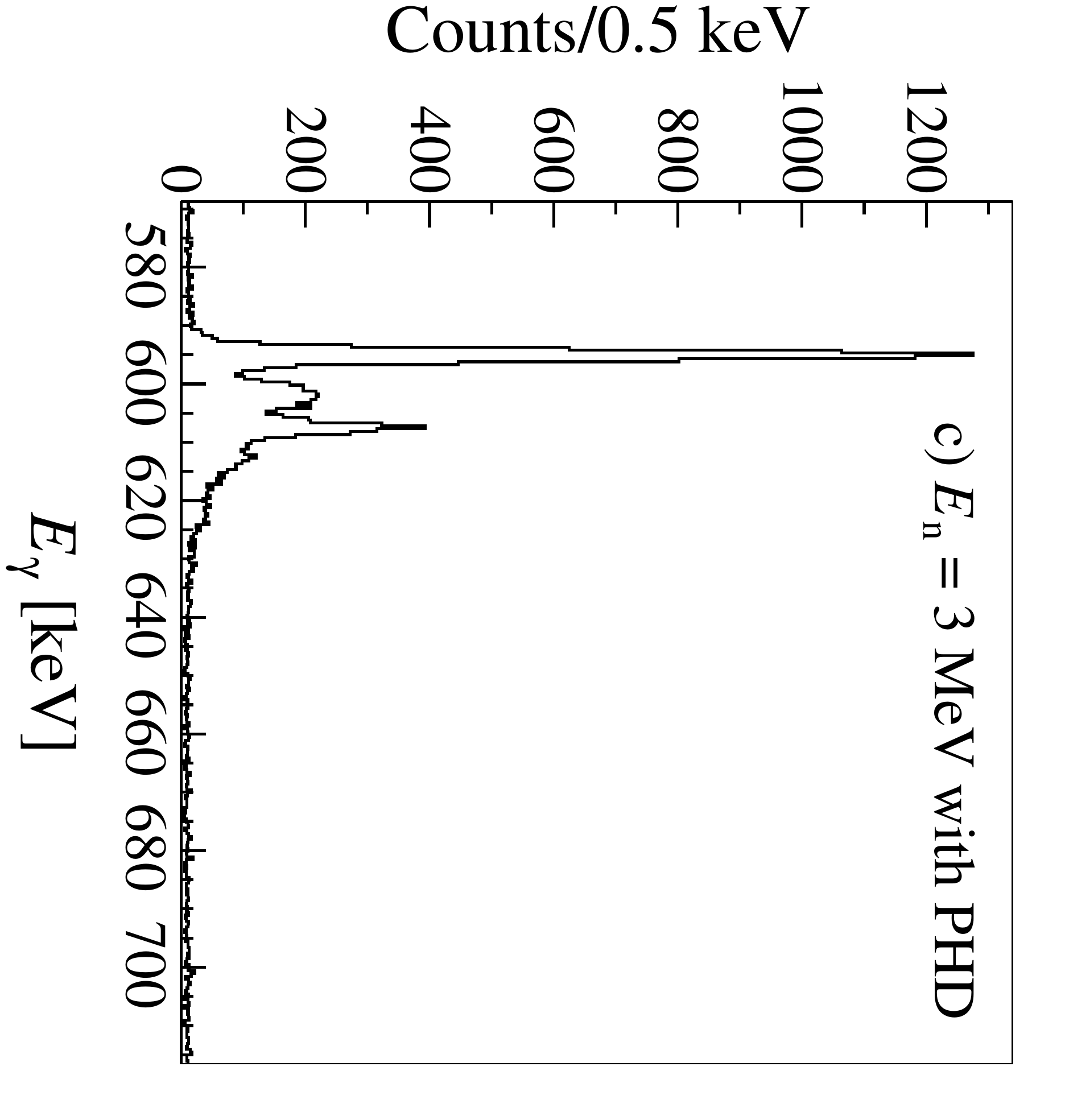}
    \includegraphics[width=0.47\columnwidth,angle=90]{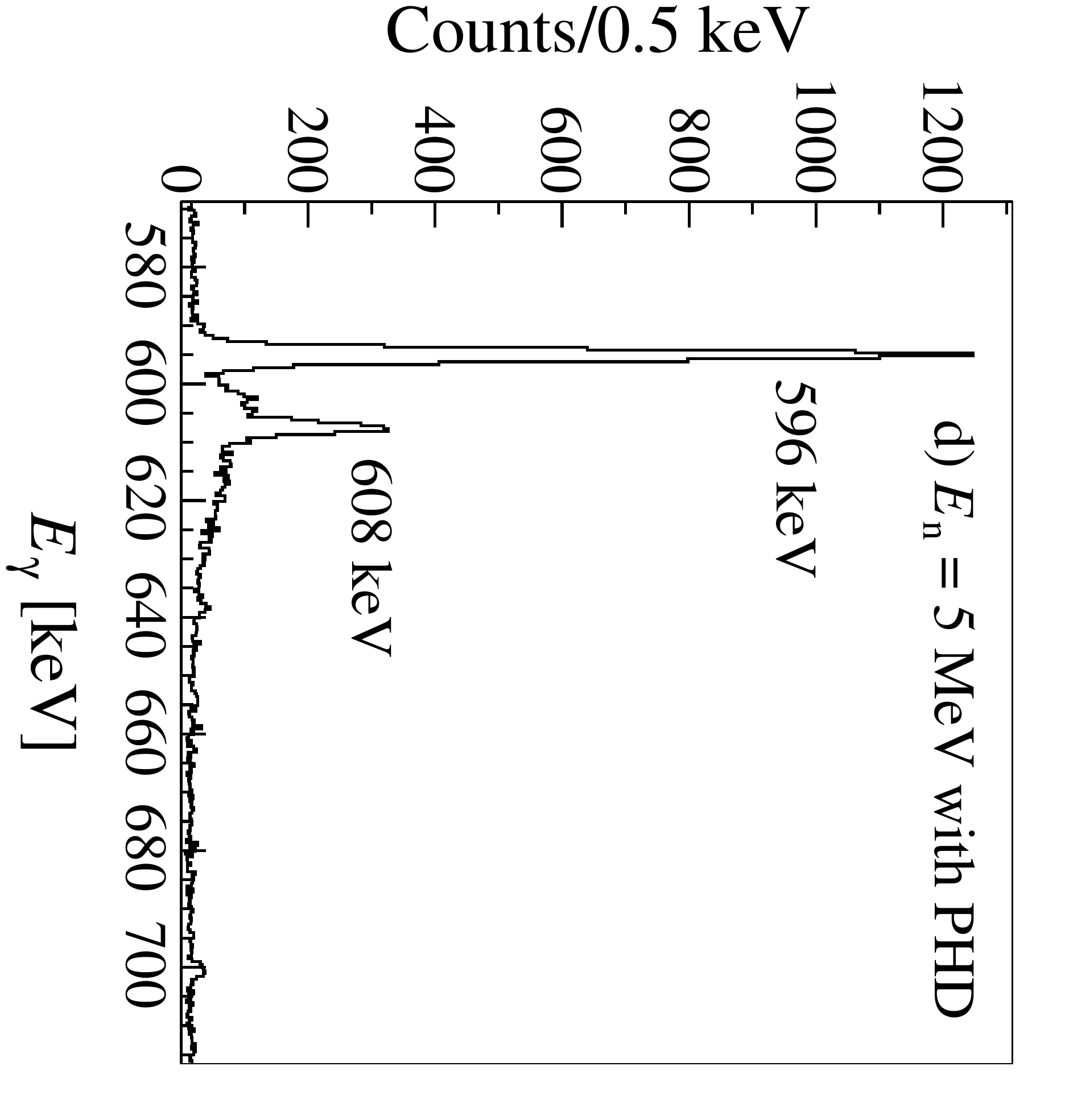}
  \end{center}
  \caption{Tracked $\gamma$-ray spectra obtained after inelastic
    scattering of neutrons on $^{74}$Ge (data sets 4 and 5).  The 596
    keV and 608 keV peaks are due to the $2_1^+ \rightarrow 0_1^+$ and
    $2_2^+ \rightarrow 2_1^+$ transitions in $^{74}$Ge, respectively.
  }
  \label{fig:09}
\end{figure}

In figs. \ref{fig:09}a) and b) the effect of the PHD correction on the
$\gamma$-ray spectra obtained by inelastic scattering of 1 MeV
neutrons on $^{74}$Ge is shown (data sets 4 and 5). Apart from the
reduced energies one can also observe a reduction of the bump and a
corresponding increase of the counts in the 596 keV full-energy peak.
This is due to the fact that some counts are moved from the bump to
the peak if the recoil energy is not measured. As shown in
figs. \ref{fig:09}b), c) and d), the bump gets wider and moves to
higher energies as the neutron energy increases from 1 MeV to 5 MeV
(cf. the recoil energy distributions shown in fig. \ref{fig:01}). When
the bump moves to higher energies less of the counts in it will be
lost due to the 5 keV energy threshold.

The PHD correction was applied in all simulations presented in the
rest of this section.

\subsection{1 MeV \texorpdfstring{$\gamma$} rays and 1-5 MeV neutrons
  (data sets 2 and 6)} \label{ss:flat}

In this subsection the results of inelastic scattering of 1-5 MeV
neutrons on $^{\textrm{nat}}$Ge are presented.  The neutrons had a
flat energy distribution and the PHD correction was applied (data set
6). The different neutron-$\gamma$ discrimination methods described in
subsection \ref{ss:methods} were tested and their effect on the
photopeak efficiency was checked by a simulation of 1 MeV $\gamma$
rays emitted from the center of AGATA (data set 2).

\begin{figure}[htb!] 
  \centering
  \begin{center}
    \includegraphics[width=0.6\columnwidth,angle=90]{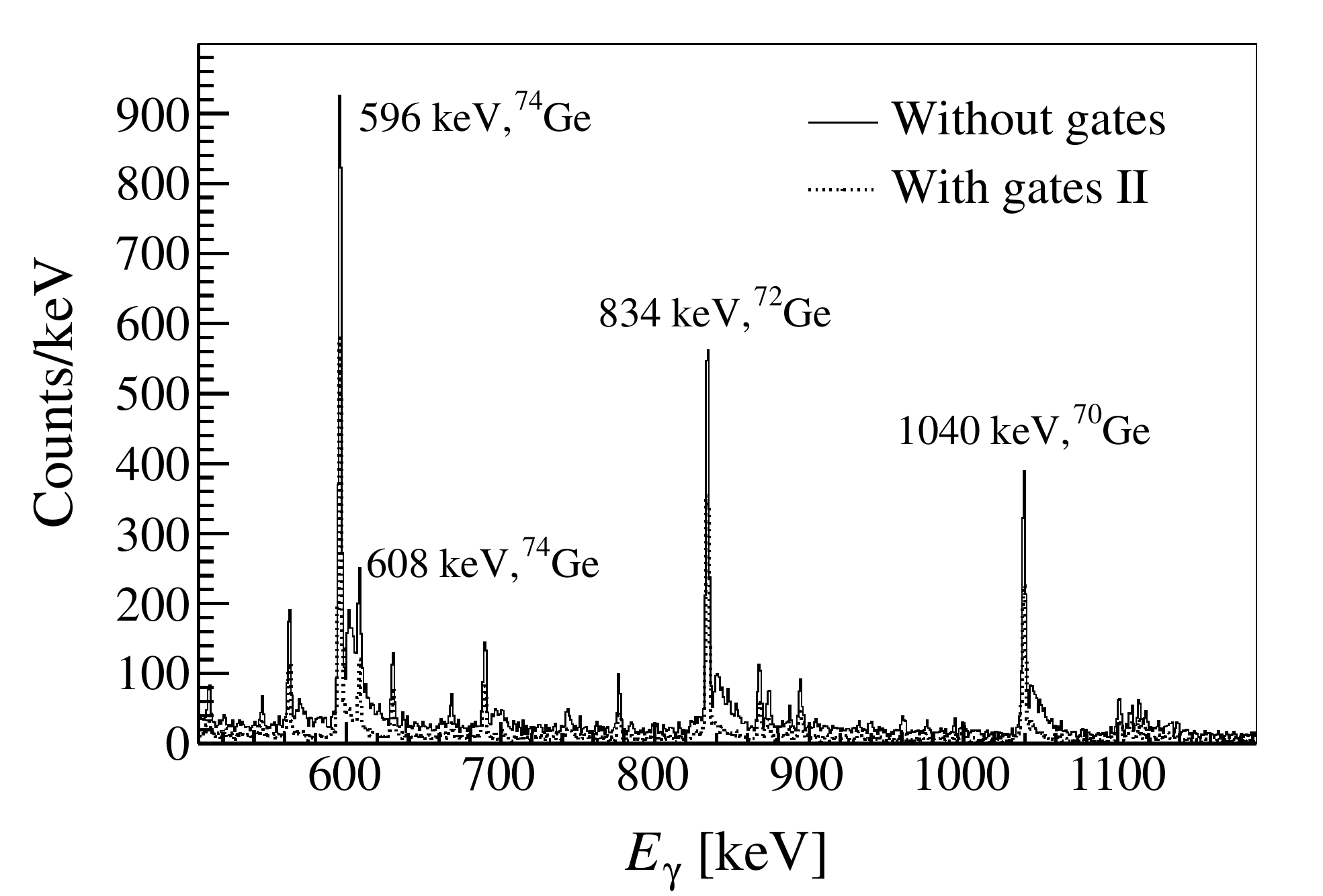}
  \end{center}
  \caption{Tracked $\gamma$-ray spectra obtained after inelastic
    scattering of 1-5 MeV neutrons on $^{\textrm{nat}}$Ge (data set
    6).  The solid histogram shows the $\gamma$-ray spectrum obtained
    without any neutron rejection gates while the dotted histogram
    shows what is left after applying gate set II ($E_1$, $E_2$,
    $\Delta\theta$, $FM$).}
  \label{fig:10}
\end{figure}

In fig. \ref{fig:10} the histogram with the solid line shows an
ungated $\gamma$-ray spectrum (no neutron rejection gates were
applied) whereas the dotted histogram shows what is left after the
application of gate set II ($E_1$, $E_2$, $\Delta\theta$, $FM$).
Quantitative results of the application of the different gates are
shown in table \ref{tab:03}.

With gate set II the total counts in the neutron induced $\gamma$-ray
spectrum was reduced by 54 \%, the 834 keV peak by 39 \%, and its bump
by 76 \% (see table \ref{tab:03}). This causes a loss of 14 \% of the
1 MeV peak in the spectrum obtained by sending 1 MeV $\gamma$ rays
from the center of AGATA (data set 2).  If instead, gate set I is
applied, the total neutron induced spectrum is reduced by 40 \%, the
834 keV peak by 23 \%, and its bump by 58 \% whereas the loss in the 1
MeV full-energy peak is only 5 \% in the 1 MeV $\gamma$-ray induced
spectrum.
\begin{table*}[htb!]
  \renewcommand{\arraystretch}{1.2}
  \caption{Reduction of counts in percent of the 834 keV peak ($2_1^+
    \rightarrow 0_1^+$ transition in $^{72}$Ge), its bump, and of the
    total spectrum after applying different neutron-$\gamma$
    discrimination gates.  The results were obtained by sending
    neutrons with a flat energy distribution from 1 to 5 MeV (data set
    6) and 1 MeV $\gamma$ rays (data set 2) on $^{\textrm{nat}}$Ge.
    The gates used were $E_1< 20$ keV, $E_2 < 15$ keV, $\Delta\theta >
    15^{\circ}$, $0.05 < FM < 1$.  The gate combinations ($E_1$, $FM$)
    and ($E_1$, $E_2$, $\Delta\theta$, $FM$) are labelled I and II,
    respectively, and were created by using a logical or of the
    individual gates.}
  \label{tab:03}
  \begin{tabular*}{\textwidth}{@{\extracolsep{\fill}}lcccc}
    \hline
    \multirow{3}{*}{Gates} 
    & \multicolumn{3}{c}{1-5 MeV neutrons (data set 6)}
    & 1 MeV $\gamma$ rays (data set 2)\\
    \cline{2-5} & 834 keV peak & Bump & Total & Peak \\
    \hline
    $E_1$
    & 1 & 38 & 11 & 1 \\
    $\Delta\theta$
    & 25 & 51 & 36 & 11 \\
    $FM$
    & 23 & 45 & 36 & 4 \\
    I: ($E_1$, $FM$) 
    & 23 & 58 & 40 & 5  \\
    II: ($E_1$, $E_2$, $\Delta\theta$, $FM$)
    & 39 & 76 & 54 & 14 \\
    \hline
  \end{tabular*}
\end{table*}

\subsection{Flat energy distribution of 1 to 5 MeV neutrons and a 
  \texorpdfstring{$\gamma$}-ray cascade (data sets 3 and 7)}
\label{ss:cascade}

\begin{figure}[htb!]
  \centering
  \includegraphics[width=0.65\columnwidth,angle=90]{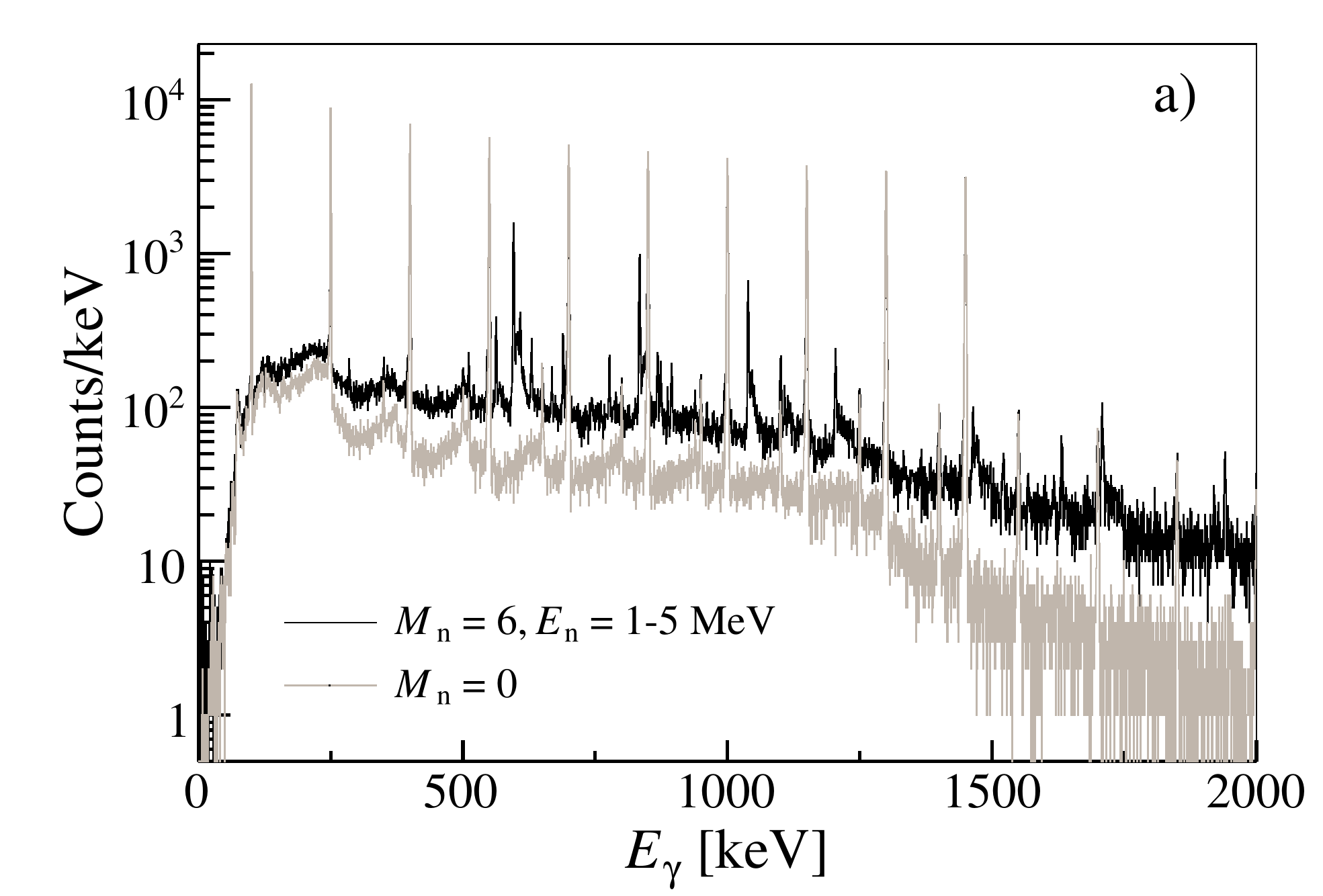} 
  \includegraphics[width=0.65\columnwidth,angle=90]{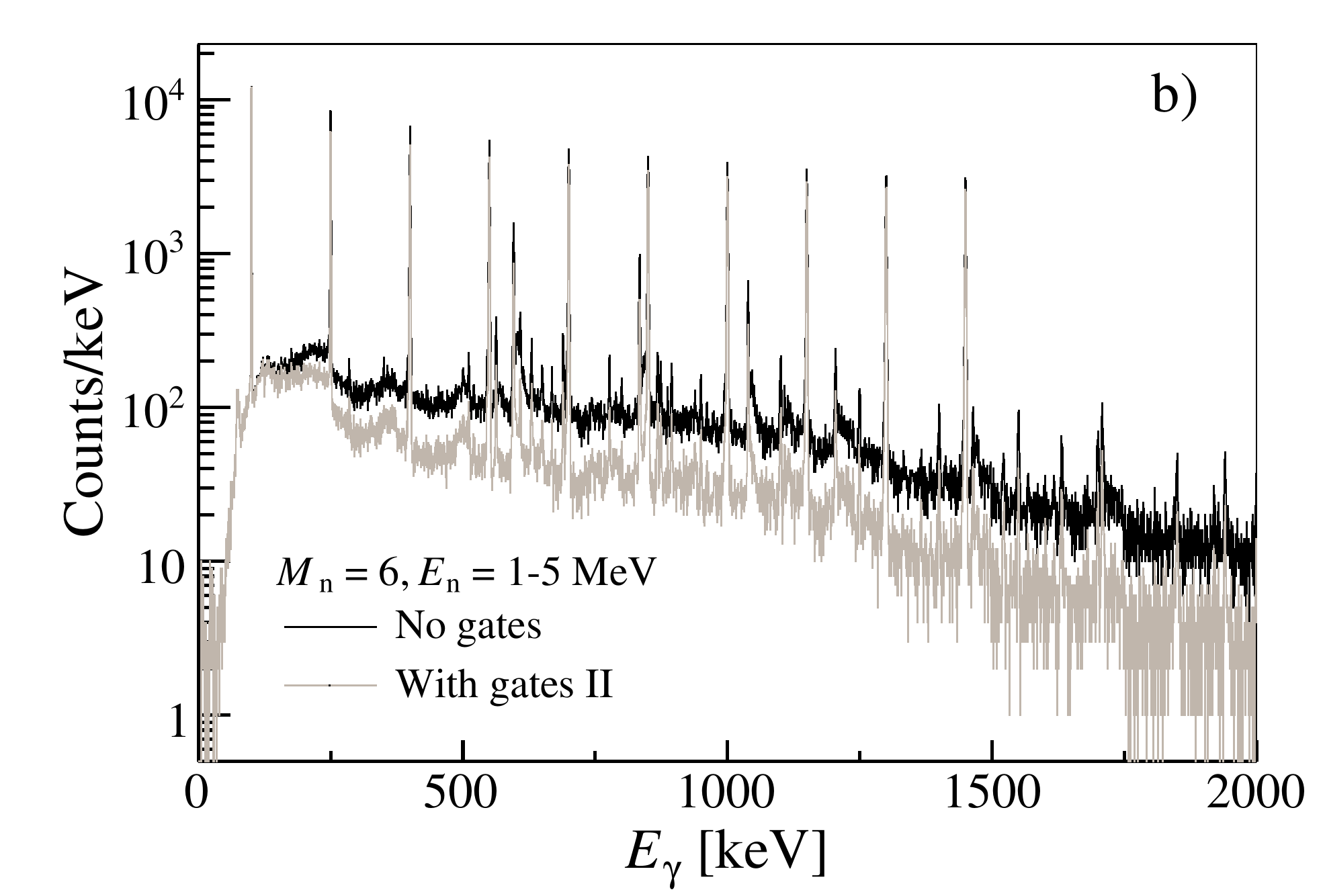} 
  \caption{a) Tracked spectrum for a $\gamma$-ray cascade of
    multiplicity 10 when no neutrons are present in the events (gray,
    data set 3) and when 6 neutrons are emitted in coincidence with
    the $\gamma$ rays (black, data set 7).  b) Tracked energy of the
    same $\gamma$-ray cascade of multiplicity 10 emitted in
    coincidence with 6 neutrons before (black) and after (gray)
    neutron rejection.}
  \label{fig:11}
\end{figure}

In this subsection the results of applying the neutron rejection
methods on events in which neutrons are emitted in coincidence with
$\gamma$ rays are presented. A simulation of simultaneous emission of
6 neutrons and a $\gamma$-ray cascade of multiplicity 10 (data set 7)
was compared to one with no neutron emission (data set 3). The
aluminium capsules of the HPGe detectors were included in this
simulation. Fig. \ref{fig:11}a shows the large and complex background
caused by the inelastic scattering of neutrons compared to the
spectrum for which no neutrons were emitted.

\begin{figure}[htb!] 
  \centering
  \begin{center}
    \includegraphics[width=0.47\columnwidth,angle=90]{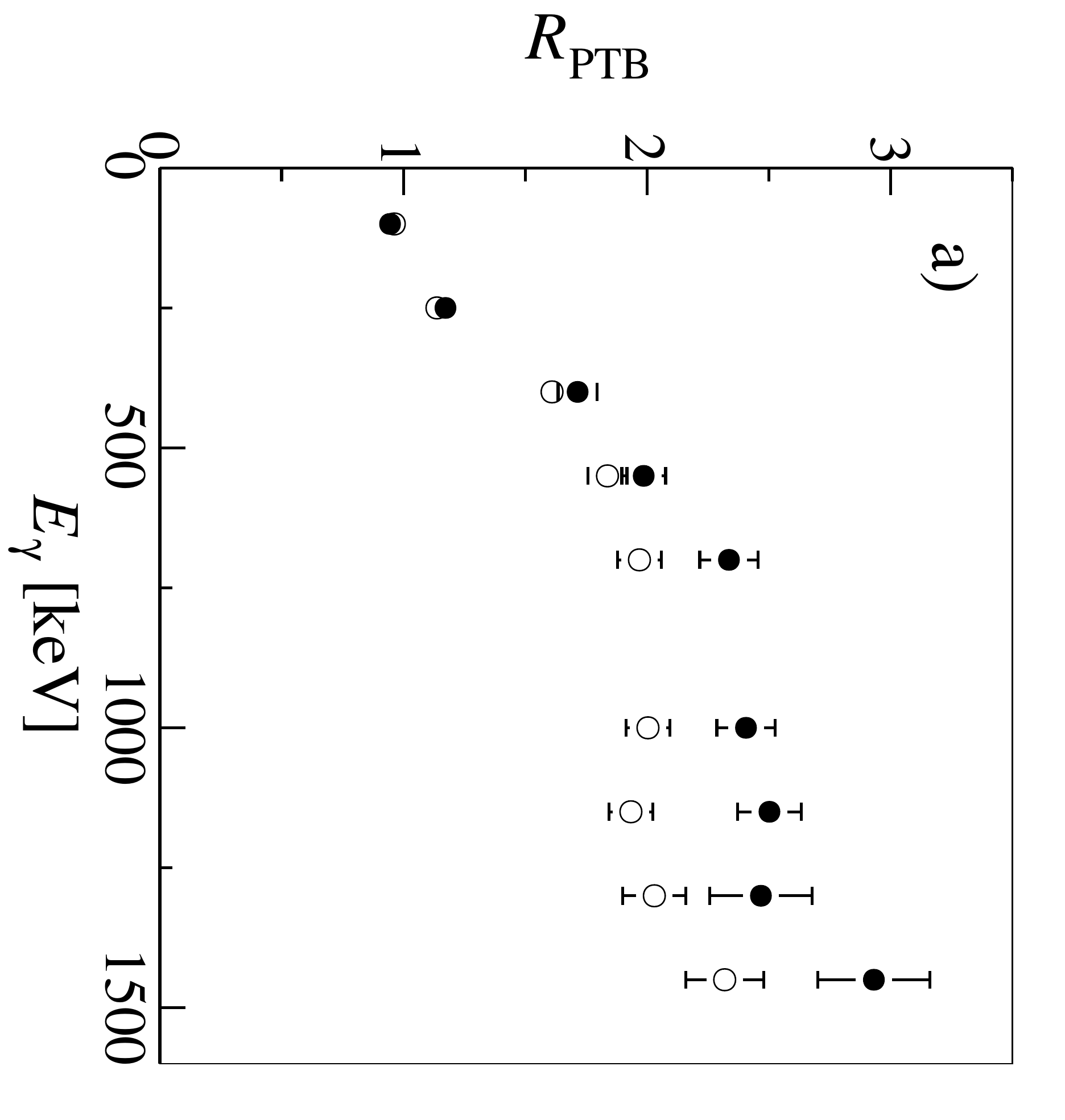}
    \includegraphics[width=0.47\columnwidth,angle=90]{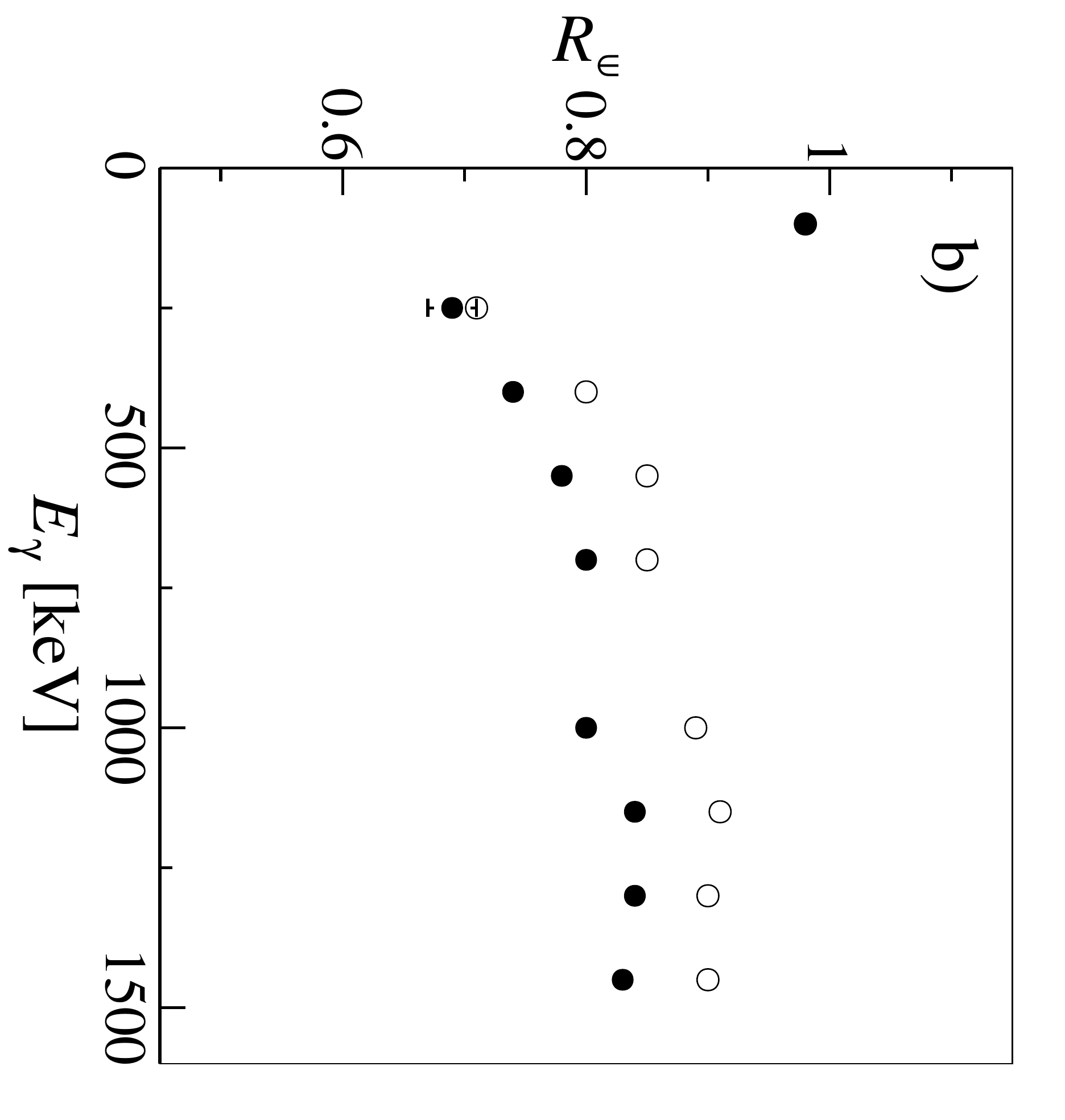}
  \end{center}
  \caption{Ratio of a) peak-to-background values $R_{\textrm{PTB}}$
    and b) photopeak efficiencies $R_{\epsilon}$ with and without
    neutron rejection for $M_{\textrm{n}} = 6$.  Open and filled
    circles were obtained by using gate set I and II, respectively.
    The data point for the 850 keV $\gamma$ ray
      was excluded from the plots because it is located on the bump of the
      834 keV transition (see fig. \ref{fig:10}) and therefore difficult to
      analyze.}
  \label{fig:12}
\end{figure}

The influence of the neutrons can be quantified by the
peak-to-background ratio (PTB) and the total photopeak efficiency
($\epsilon$) of the $\gamma$-ray peaks in the spectra. The results are
summarized in tables \ref{tab:04} and \ref{tab:05}. The PTB ratio for
a $\gamma$-ray peak was determined as the number of counts in the peak
area divided by the background area, which was defined as a region of
$\pm 2 \sigma$ around the centroid of the $\gamma$-ray peak, with
$\sigma$ being the width of the peak.  The two gate sets, I and II,
were are used for the neutron-$\gamma$ discrimination.  The ratios of
the PTB values, $R_{\textrm{PTB}}$, which show the improvement of the
PTB values when applying the neutron rejection gates I and II on the
data set with $M_{\textrm{n}} = 6$, are also given in the table. The
ratios $R_{\textrm{PTB}}$ are also plotted as a function of the
$\gamma$-ray energy in fig. \ref{fig:12}a.
\begin{table*}[htb!]
  \renewcommand{\arraystretch}{1.2}
  \caption{Peak-to-background ratios (PTB) for the emission of 0 and 6
    neutrons in coincidence with a $\gamma$-ray cascade of
    multiplicity 10 (data sets 2 and 7). The $R_{\textrm{PTB}}$ values
    are ratios of the PTB ratios extracted from the gated and ungated
    spectra with $M_{\textrm{n}} = 6$ using gate set I (column
    4/column 3) and II (column 5/column 3).  See caption of table
    \ref{tab:03} for a description of gates I and II.  }
  \label{tab:04}
  \begin{tabular*}{\textwidth}{@{\extracolsep{\fill}}ccccccc}
    \hline
    \multirow{3}{*}{$E_{\gamma}$} & 
    \multicolumn{4}{c}{Peak-to-background ratio, PTB} &
    \multicolumn{2}{c}{$R_{\textrm{PTB}}$} \\
    \cline{2-7}
    & No gates & No gates & Gates I & Gates II  & 
    {\multirow{2}{*}{I}} & {\multirow{2}{*}{II}} \\
    \cline{2-5}
    & $M_{\textrm{n}} = 0$ &  $M_{\textrm{n}} = 6$ & $M_{\textrm{n}} = 6$
    & $M_{\textrm{n}} = 6$ & & \\
    \hline
     100  &  32.0 & 25.2 & 23.3 & 23.6 & 0.93 & 0.94 \\
     250  &  16.6 & 10.9 & 12.0 & 12.8 & 1.10 & 1.17 \\
     400  &  34.6 & 15.5 & 24.3 & 26.5 & 1.57 & 1.71 \\
     550  &  36.5 & 15.3 & 27.6 & 30.3 & 1.80 & 1.98 \\
     700  &  35.7 & 15.4 & 29.7 & 35.9 & 1.93 & 2.33 \\
     850  &  35.8 &  8.3 & 17.9 & 22.1 & 2.15 & 2.65 \\
    1000  &  37.5 & 13.7 & 26.9 & 32.8 & 1.97 & 2.40 \\
    1150  &  35.4 & 13.7 & 26.0 & 34.3 & 1.89 & 2.50 \\
    1300  &  57.2 & 18.7 & 40.1 & 49.5 & 2.00 & 2.46 \\
    1450  & 110.2 & 25.0 & 56.9 & 73.2 & 2.28 & 2.93 \\
    \hline
  \end{tabular*}
\end{table*}

As seen in figs. \ref{fig:11}a and b, the neutron emission as well as
the neutron rejection method has a major influence on the PTB ratios.
On the average, the emission of 6 neutrons reduces the PTB ratio by a
factor of 2.6, whereas the neutron rejection method increases it by a
factor 1.8 and 2.1 for gates I and II, respectively. At a $\gamma$-ray
energy of 1 MeV, the PTB ratio is improved from 13.7 to 32.8, i.e. by
a factor of 2.4 for gates II. The PTB ratio at 1 MeV is 37.5 for a
simulation with no emitted neutrons ($M_{\textrm{n}} = 0$).

The influence of the neutron emission and of the neutron rejection
methods on the photopeak efficiency was also investigated. The
emission of neutrons have a negligible influence on the photopeak
efficiency. This can be seen by comparing the $\epsilon$ values for
$M_{\textrm{n}} = 6$ to the once obtained for $M_{\textrm{n}} = 0$ in
table \ref{tab:05}.  The neutron rejection, however, decreases the
$\epsilon$ values. On the average, $\epsilon$ is reduced by a factor
of 1.16 and 1.24 for the neutron rejection gates I and II,
respectively. The $R_{\epsilon}$ values plotted in fig. \ref{fig:12}b
are the ratios of the $\epsilon$ values with and without neutron
rejection for gates I and II and $M_{\textrm{n}} = 6$.  

By using gate set II, the photopeak efficiency of the $\gamma$-ray
peak at 1 MeV was reduced by a factor of 1.25 (20 \%) while the
$\epsilon$ value of the 1039 keV peak, due to the $2_1^+ \rightarrow
0_1^+$ transition in $^{70}$Ge, was reduced by a factor of 2.02 (51
\%).  With gate set I the $\epsilon$ value of the $\gamma$-ray peak at
1 MeV was reduced by a factor of 1.13 (11 \%) and the 1039 keV by a
factor of 1.57 (36 \%).

A test of the dependency of the $R_{\textrm{PTB}}$ and $R_{\epsilon}$
ratios on the low-energy threshold was also performed. At a
$\gamma$-ray energy of 1 MeV and with gates II the $R_{\textrm{PTB}}$
was reduced from 2.4 with an energy threshold of 5 keV to 1.9 with a
threshold of 30 keV.  The $R_{\epsilon}$ value was not effected by the
energy threshold.
\begin{table*}[htb!]
  \renewcommand{\arraystretch}{1.2}
  \caption{Photopeak efficiency ($\epsilon$) for the emission of 0 and
    6 neutrons in coincidence with a $\gamma$-ray cascade of
    multiplicity 10 (data sets 2 and 7). The $R_{\epsilon}$ values are
    ratios of the photopeak efficiencies extracted from the gated and
    ungated spectra with $M_{\textrm{n}} = 6$ using gate set I (column
    4/column 3) and II (column 5/column 3). See caption of table
    \ref{tab:03} for a description of gates I and II.}
  \label{tab:05}
  \begin{tabular*}{\textwidth}{@{\extracolsep{\fill}}ccccccc}
    \hline
    \multirow{3}{*}{$E_{\gamma}$} & 
    \multicolumn{4}{c}{Photopeak efficiency, $\epsilon$} &
    \multicolumn{2}{c}{$R_{\epsilon}$} \\
    \cline{2-7}
    & No gates & No gates & Gates I & Gates II  & 
    {\multirow{2}{*}{I}} & {\multirow{2}{*}{II}} \\
    \cline{2-5}
    & $M_{\textrm{n}} = 0$ &  $M_{\textrm{n}} = 6$ & $M_{\textrm{n}} = 6$
    & $M_{\textrm{n}} = 6$ & & \\
    \hline
    100   & 0.60 & 0.57 & 0.56 & 0.56 & 0.98 & 0.98 \\
    250   & 0.52 & 0.49 & 0.35 & 0.34 & 0.71 & 0.69 \\
    400   & 0.48 & 0.46 & 0.37 & 0.34 & 0.80 & 0.74 \\
    550   & 0.44 & 0.41 & 0.35 & 0.32 & 0.85 & 0.78 \\
    700   & 0.41 & 0.39 & 0.33 & 0.31 & 0.85 & 0.80 \\
    850   & 0.38 & 0.36 & 0.32 & 0.29 & 0.89 & 0.81 \\
    1000  & 0.37 & 0.35 & 0.31 & 0.28 & 0.89 & 0.80 \\
    1150  & 0.35 & 0.32 & 0.29 & 0.27 & 0.91 & 0.84 \\
    1300  & 0.23 & 0.31 & 0.28 & 0.26 & 0.90 & 0.84 \\
    1450  & 0.31 & 0.30 & 0.27 & 0.25 & 0.90 & 0.83 \\
    \hline
  \end{tabular*}
\end{table*}

\section{Summary and conclusions} \label{sec:concul}

Monte Carlo simulations of neutrons interacting with the HPGe
detectors of the AGATA spectrometer were performed. The possibility of
using $\gamma$-ray tracking for discrimination of interactions due to
inelastic scattering of neutrons on Ge was investigated with the aim
to suppress the neutron induced background in the $\gamma$-ray
spectra. Three methods were developed to find ``fingerprints'' of the
neutron interaction points in the detectors and to use them for
neutron rejection.  The first method makes use of the low Ge recoil
energies deposited in the detectors by the scattered neutrons. The
second method utilizes the random incoming direction of the initial
$\gamma$ rays that are produced by inelastic neutron scattering. The
third method makes use of differences in the figure-of-merit values
evaluated by the $\gamma$-ray tracking program for $\gamma$-ray and
neutron induced interactions.  These methods were first tested on a
simplified case in which neutrons with energies 1 to 5 MeV and
multiplicity 1 were emitted from the center of the AGATA
spectrometer. With a combination of the three neutron-$\gamma$
discrimination methods, the 834 keV transition in $^{72}$Ge was
reduced by 39 \% whereas the bump structure that belongs to this
transition was reduced more effectively, by 76 \%. The bump originates
from the tracking of clusters where the Ge recoil energy is added to
the energy of the $\gamma$ ray produced by inelastic scattering of the
neutron. The neutron rejection methods were also tested on a case
where 6 neutrons with energies between 1 to 5 MeV were emitted in
coincidence with a cascade of 10 $\gamma$ rays. The influence of the
neutron emission and the neutron rejection methods were quantified by
determining the peak-to-background ratios and total photopeak
efficiencies of the $\gamma$-ray peaks in the tracked spectra. As a
result, two sets of gates, which are combinations of the three neutron
rejection methods, were suggested and used.  At a $\gamma$-ray energy
of 1 MeV, the first gate set increases the peak-to-background value by
a factor of 2.0, whereas the second set of gates increases it by a
factor of 2.4. The same gates reduce the photopeak efficiency at 1 MeV
by a factor of 1.13 and 1.25, respectively. The future plan is to test
the neutron-$\gamma$ discrimination methods on real data obtained by
the AGATA spectrometer. Neutron-$\gamma$ discrimination is also
important for measurements of high-energy $\gamma$ rays e.g. from the
decay of giant resonances at high temperatures \cite{2006PRL.97}.  For
this purpose it is necessary to investigate neutron interactions in
the energy interval between 5 to 20 MeV. This work is ongoing.

\section*{Acknowledgments}

We are grateful to D. Bazzacco and E. Farnea for the {\mgt} and
AGATA {\geant} codes. This work was supported by the Scientific and
Technological Council of Turkey (Proj. no. 106T055), Ankara University
(BAP Proj. no. 05B4240002), EURONS AGATA (Contract no. 506065-R113)
and the Swedish Research Council.

\end{document}